\documentclass[preprint]{revtex4-2}
\usepackage{graphicx} 
\usepackage[colorlinks=true,linkcolor=blue,citecolor=blue,urlcolor=black]{hyperref}
\usepackage[english]{babel}
\usepackage{datetime}
\usepackage{verbatim}
\usepackage{amsmath}
\usepackage{amssymb}
\usepackage{siunitx} 
\usepackage{nccmath}
\usepackage{dcolumn}
\usepackage{lineno}
\usepackage{xcolor}
\usepackage{color}
\usepackage{float}
\usepackage[utf8]{inputenc}
\usepackage{lineno}
\usepackage[symbolgreek]{mathastext}
\usepackage{import}

\begin{document}

\widetext
\raggedbottom

\title{\Large Self-reconfiguring colloidal active matter}

\author{Stefania Ketzetzi}
\affiliation{Laboratory for Soft Materials and Interfaces, Department of Materials, ETH Zürich, Switzerland.}
\affiliation{Present address: John A. Paulson School of Engineering and Applied Sciences, Harvard University, Cambridge, MA 02138, USA}
\author{Lorenzo Caprini}
\affiliation{Institut f\"{u}r Theoretische Physik II: Weiche Materie, Heinrich-Heine-Universit\"{a}t D\"{u}sseldorf, D-40225 D\"{u}sseldorf, Germany}
\affiliation{Present address: Physics Department, Universit\'a di Roma La Sapienza, P.le Aldo Moro 2, 00185, Rome, Italy}
\author{Vivien Willems}
\affiliation{Laboratory for Soft Materials and Interfaces, Department of Materials, ETH Zürich, Switzerland.}
\affiliation{Present address: University of Bordeaux, CNRS, CRPP, UMR 5031, F-33600 Pessac, France}
\author{Laura Alvarez}
\affiliation{Laboratory for Soft Materials and Interfaces, Department of Materials, ETH Zürich, Switzerland.}
\affiliation{Present address: University of Bordeaux, CNRS, CRPP, UMR 5031, F-33600 Pessac, France}
\author{Hartmut L\"{o}wen}
\affiliation{Institut f\"{u}r Theoretische Physik II: Weiche Materie, Heinrich-Heine-Universit\"{a}t D\"{u}sseldorf, D-40225 D\"{u}sseldorf, Germany}
\author{Lucio Isa}
\email{Corresponding author: lucio.isa@mat.ethz.ch}
\affiliation{Laboratory for Soft Materials and Interfaces, Department of Materials, ETH Zürich, Switzerland.}

\date{\today}

\begin{abstract}

Cells and microorganisms employ dynamic shape changes to enable steering and avoidance for efficient spatial exploration and collective organization. In contrast, active colloids, their synthetic counterparts, currently lack similar abilities and strategies. Through physical interactions alone, here we create active colloidal molecules that spontaneously reconfigure their structure, unlike traditional active particles. We find that self-reconfiguration decouples reorientational dynamics from rotational diffusivity and bestows our active molecules additional reorientation capabilities. During encounters with neighbors, rapid conformational changes lead to self-steering and avoidance. At higher area fractions, reconfiguration-induced avoidance fully inhibits characteristic dynamic clustering, motility-induced phase separation and flocking; instead, the system retains a homogeneous structure comprising well-separated active units. Self-reconfiguring systems therefore present an exciting path towards autonomous motion beyond that of classical synthetic active matter.

\end{abstract}

\maketitle

\subsection*{\large Introduction}
A defining goal of soft matter research is to recapitulate the unique breadth of biological behaviors and functions in synthetic materials. To that end, synthetic microswimmers, or active colloidal particles, are often proposed as minimal models displaying physical behavior inspired by biological systems across length scales~\cite{Dreyfus2005,Bechinger2016,Aubret2018,Muinos2021,Zottl2016}. As all living organisms~\cite{Buhl2006,Flack2018,Charlesworth2019}, cells and microorganisms steer themselves through dynamic shape regulation, sensing, communication and information exchange with their surroundings. Dynamic reshaping and spontaneous reconfiguration enhance spatial exploration capabilities and promote collective organization that contribute to ensuring survival across different environments. Aligning with neighbors, for instance, promotes coordinated migration resulting in swarming dynamics~\cite{Kearns2010,Schaller2010, Sanchez2012,Sumino2012, Henkes2020}, while anti-aligning interactions promote self-avoidance, i.e. a change in the direction of motion upon encountering a neighbor, resulting in spreading. Both autonomous steering and avoidance are broadly leveraged across cell biology, e.g., during development and repair, to optimize efficiency and ensure proper dispersion~\cite{Stramer2017,Singh2021, Lammermann2008, Carmona2008}. 

However, to date, those traits still escape experimental realization in synthetic microswimmers. Indeed, active colloids are mechanically rigid and pre-configured units, lacking both flexibility and self-reconfiguration as well as active response to their surroundings. As a consequence, current realizations cannot fully reproduce the autonomous organization and steering behavior typical of biological units. Translation and rotation are coupled through the temperature of the bath, connecting motion persistence to rotational diffusion~\cite{Bechinger2016}, different from biological self-steering. A constant self-propulsion velocity and motion persistence, together with excluded volume interactions, lead to dynamic clustering, motility-induced phase separation, and other types of coordinated organizations, such as swarms, chains, and vortices~\cite{Palacci2013,Theurkauff2012,Buttinoni2013,Ginot2018,Bricard2013,Digregorio2018,Bricard2015,Yan2016,Kaiser2017,Karani2019,Soni2019,Linden2019,Geyer2019,Ketzetzi2022,Zion2022}. Recently, it has been shown that avoidance due to purely repulsive interactions leads rigid particles to flock and to form active Wigner crystals~\cite{Das2024}. Incorporating internal reconfiguration in synthetic active matter may thus allow novel, flexible modes of motion and organization on both the single and collective level.

\begin{figure*}[hb!]
    \centering
     \includegraphics[width=1\linewidth]{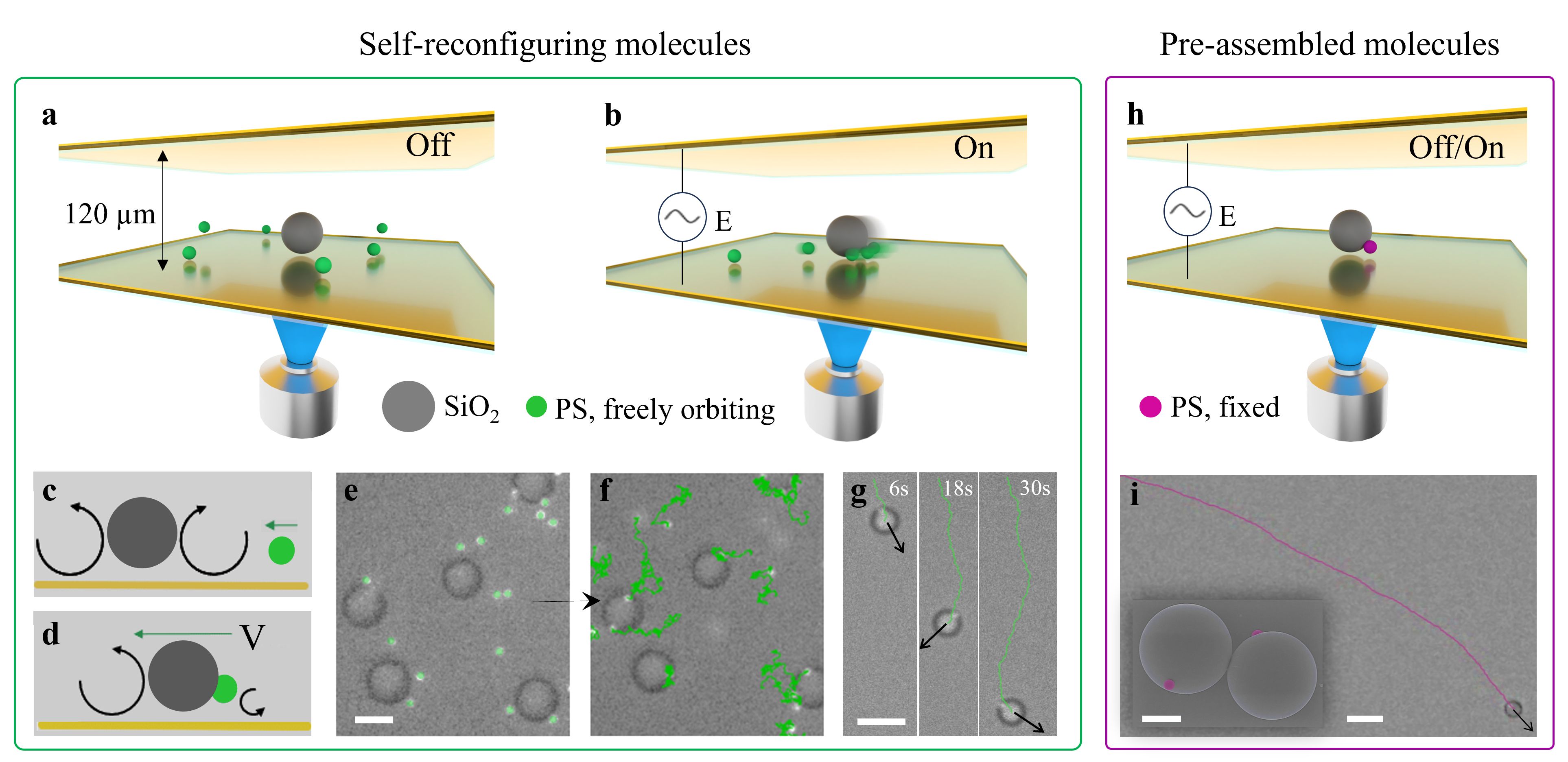}
    \vspace{-15pt}
    \caption{\textbf{Self-reconfiguring (a-g) and pre-assembled (h-i) active colloidal molecules.} \textbf{a-b,} Schematic representation of the experimental system. \textbf{a,} SiO$_2$ and polystyrene (PS) colloids suspended in water are confined between two transparent conductive surfaces for observation with an inverted microscope. \textbf{b,} Under a perpendicular AC electric field, colloids self-assemble into self-reconfiguring active molecules. \textbf{c-d,} Schematics of the assembly and propulsion mechanism. \textbf{c,} At a frequency of 1 kHz, electrohydrodynamic flows along the SiO$_2$ surfaces attract PS particles in their vicinity. \textbf{d,} Upon increasing AC voltage peak-to-peak amplitude ($V_{pp}$), PS colloids assemble around the SiO$_2$, causing an asymmetry in the fluid flow that propels the molecules forward. \textbf{e-g,} Optical micrographs depicting molecule formation and motion (PS and their trajectories in green). \textbf{e,} At 1 $V_{pp}$, PS colloids approach the SiO$_2$ (scale bar 5 \textmu m). \textbf{f,} At 2 $V_{pp}$, PS colloids (satellites) assemble around the SiO$_2$ (core) forming molecules. The PS satellites continuously translate along the cores' surface. \textbf{g,} Active trajectory of a reconfiguring dimer at different times (6 $V_{pp}$, scale bar 10 \textmu m). \textbf{h,} Schematic of control experiments featuring pre-assembled molecules made from the same colloids, albeit with the PS fixed on the SiO$_2$ surface. \textbf{i,} Active trajectory of a pre-assembled dimer (in magenta). Inset: SEM image of two pre-assembled dimers (PS colloids here in magenta). Scale bars are 10 and 2 \textmu m, respectively.}
    \label{fig:fig1}
\end{figure*}

Here, we realize synthetic microswimmers in the form of active colloidal molecules that self-assemble, move, and self-reconfigure based on physical interactions. We find that these molecules dynamically steer their direction of motion and actively avoid neighbors via effective anti-alignment interactions. Reconfiguration-induced avoidance fully prevents the aforementioned dynamic clustering, motility-induced phase separation and flocking observed in (repulsive) rigid active particles, and instead enables obtaining densely-packed homogeneous systems of well-separated actively-moving units. Through experiments and simulations, we show that reconfiguration decouples the reorientation of the molecule from its rotational diffusivity, shifting the timescale for motion persistence to that for internal reshaping. These features, highly desirable for ``intelligent" synthetic microswimmers, are clearly distinct from those exhibited by traditional mechanically pre-configured active colloids and pave the way towards exploring complex biological behaviors with synthetic particles~\cite{Cleland2009, Huang2021,Angelini2011,Kaiser2018}.

\subsection*{\large Self-reconfiguring active molecules}
We trigger the formation and self-propulsion of active colloidal molecules without any chemical fuel using an alternating current AC electric field~\cite{Shields2017,Zhang2016,Han2018,Alvarez2021,Wang2020,Ma2015,Ma2015PNAS} and further demonstrate that they exhibit self-reconfiguration stemming from purely physical interactions between their constituents. We mix aqueous suspensions of silica (SiO$_2$) and fluorescently-labelled polystyrene (PS) colloids that differ in size and dielectric properties (Appendix). The SiO$_2$ and PS radii are R$_c$ = 2.8 \textmu m and R$_s$ = 0.35 \textmu m, respectively. We confine the suspensions between two transparent conductive surfaces separated by a spacer (thickness $2H$ = 120 \textmu m), and connect the system to a function generator through which we apply a transverse AC electric field of $V_{pp}/2H$, where $V_{pp}$ is the peak-to-peak voltage amplitude. As $H$ is fixed in our experiments, we vary and report data as a function of $V_{pp}$. We follow all colloids with an inverted microscope at 10 frames per second under simultaneous bright-field and fluorescence illumination. 

Absent the field, colloids sediment and diffuse near the bottom surface (Fig.~\ref{fig:fig1}a). At frequency 1 kHz, SiO$_2$ colloids effectively act as dipoles that distort the electric double layer along the electrode, creating electrohydrodynamic (EHD) flows. These flows point inwards in the direction perpendicular to the electrode attracting the PS colloids in their vicinity (Fig.~\ref{fig:fig1}b to \ref{fig:fig1}e), see also Supplemental Fig. S1 for detailed calculations based on EHD theory~\cite{Alvarez2021}. Upon increasing the amplitude $V_{pp}$, nearby PS colloids assemble onto the SiO$_2$ surfaces forming colloidal molecules, with the SiO$_2$ colloids acting as the \textit{core} of the molecules and the PS particles behaving as \textit{satellites} (Fig.~\ref{fig:fig1}f and Ref.~\cite{Ma2015, Ma2015PNAS,Yang2019} for details on field-driven assembly). Upon assembly, PS colloids break the initial radial symmetry of the flow around their SiO$_2$ core, thus generating an asymmetric flow that self-propels the entire molecule with the PS colloids at its back (Fig.~\ref{fig:fig1}d and~\ref{fig:fig1}g). An example of the full process, including molecule self-assembly and propulsion under the electric field and with increasing $V_{pp}$, can be found in Movie 1.

Crucially, here the PS colloids are free to orbit around the SiO$_2$ core: driven by Brownian fluctuations, satellites continuously undergo translational diffusion along an orbit slightly below the core's equatorial plane (Fig.~\ref{fig:fig1}g). To pinpoint the effect of reconfiguration on active motion, we also pre-assemble molecules with fixed configurations using the same set of colloids (Fig.~\ref{fig:fig1}h and Appendix) but permanently attaching the PS colloids onto the silica particles. A scanning electron microscopy image of two pre-assembled dimers and a corresponding active trajectory under the same field are depicted in Fig.~\ref{fig:fig1}i.

\subsection*{\large Reconfiguration-induced reorientation}

\begin{figure*}[hb!]
    \centering
     \includegraphics[width=1\linewidth]{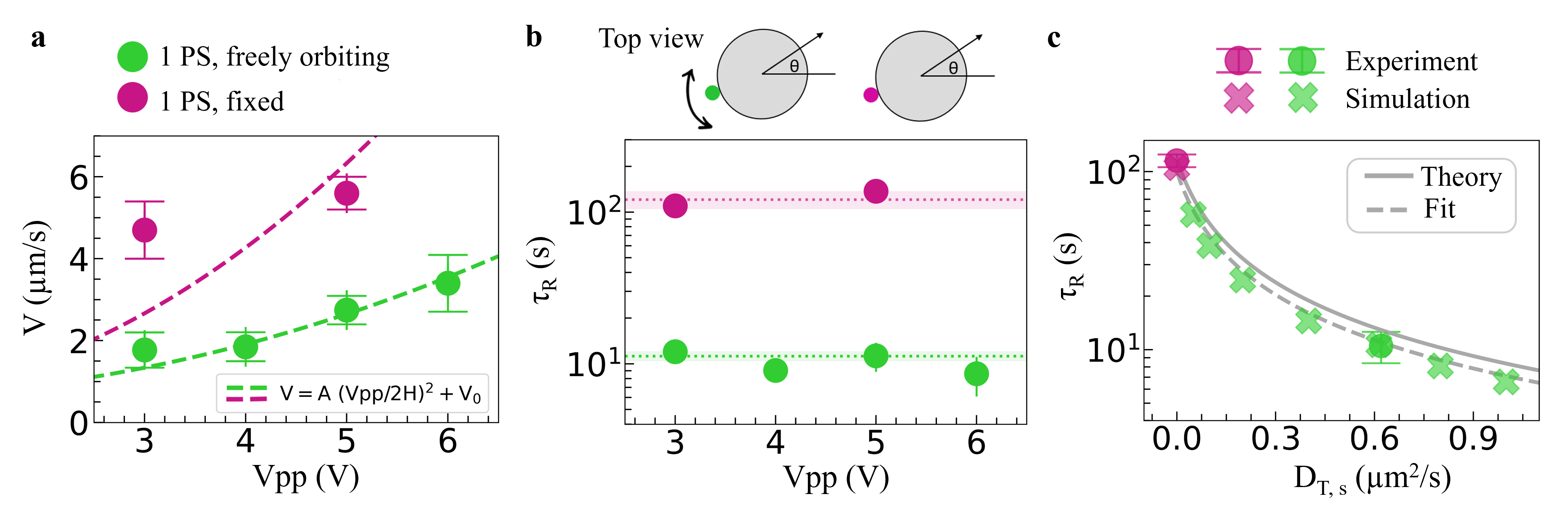}
     \vspace{-20pt}
    \caption{\textbf{Decoupling reorientation from rotational diffusivity.} \textbf{a,} Instantaneous velocity $V$ and \textbf{b,} timescale of reorientation $\tau_R$ as a function of peak-to-peak amplitude $V_{pp}$ measured for both reconfiguring and pre-assembled dimers. In a, dashed lines are fits following Ref.~\cite{Alvarez2021}; in b, dotted lines show mean values, shaded areas standard deviations. The schematic shows a top view of the dimers, with $\theta$ representing the orientation of the velocity vector. \textbf{c,} Timescale of reorientation as a function of satellite translational diffusivity $D_{T,s}$ from simulations and experiments. Errors represent standard errors.}
    \label{fig:fig2}
\end{figure*}

{\paragraph*{Molecules with minimal composition}--} First, we examine dimers, comprising a core and one satellite, as a minimal system. We track positions and quantify active motion through standard particle tracking algorithms~\cite{Trackpy}. We calculate the average speed $V$ = 
$\mid\Delta \vec{r}\mid $/$\Delta t$ of each core and hence molecule, with $\Delta t$ = 0.1 s set by the frame rate. Our measured velocities agree with EHD theory calculations that account for system-specific parameters (Supplemental Fig. S1). We find that the active velocity increases with increasing peak-to-peak voltage (Fig.~\ref{fig:fig2}a) following literature. The dotted line represents a least-squares fit with the expression $V$~=~$V_0$ + A ($V_{pp}$/2H)$^2$, where $V_0$ = (0.6 $\pm$ 0.3) \textmu m/s is the average speed of a single core particle obtained from its Brownian displacements at $\Delta t$ = 0.1 s and A is an experiment-dependent prefactor~\cite{Alvarez2021}. We verify that the configuration of the molecule with respect to the electrode remains unaffected with increasing voltage, directly from the particle positions and from the satellite-core separation distance (Supplemental Fig. S2). Likewise, the velocity of pre-assembled molecules of the same composition also increases with field amplitude, albeit with a different prefactor in line with a different orientation relative to the substrate (Supplemental Fig. S2).

We then quantify the orientational dynamics of the reconfiguring molecules by tracking the orientation angle $\theta$ of the velocity vector between frames. We extract the reorientation time $\tau_R$ of each molecule by fitting its mean square angular displacement (MSAD) \textlangle$\Delta\theta(t)^2$\textrangle~= \textlangle $\lvert\theta(t+t_0)$ - $\theta(t_0)\rvert^2$\textrangle ~as a function of time with the expression \textlangle$\Delta\theta(t)^2$\textrangle~= 2t/$\tau_R$ (Fig.~\ref{fig:fig2}b and Supplemental Fig. S2). We present the values of $\tau_R$ as a function of field amplitude for both types of molecules in Fig.~\ref{fig:fig2}b (error bars not visible in logarithmic scale). 

Strikingly, the measured timescales for reorientation differ by an order of magnitude, irrespective of field amplitude, even if the two objects have essentially the same dimensions and move under the same conditions. We find that the timescale for pre-assembled molecules $\tau_R$~=~(121 $\pm$ 16) s agrees well with the prediction for the timescale of rotation of the core t$_{R, c}$ = 1/D$_{R, c}$ = 135 s given by the Stokes--Einstein equation for rotational diffusion, even including small corrections due to presence of the wall (D$_{R, c}$ = k$_B$T / 8$\pi\eta R_c^3$ with k$_B$ the Boltzmann constant,  R$_c$ the core radius, $\eta$ and T the shear viscosity and temperature of the bath, respectively). We recover the same value for the case of two fixed PS particles (Supplemental Fig. S2). At the same time, the $\tau_R$ = (11 $\pm$ 1) s measured for the reconfiguring molecules is much faster than the Brownian rotation of the core particle. The measured reorientation time instead agrees well with the timescale for one-dimensional satellite diffusion t$_{T, s}$ $\propto$ $\rho^2 / D_{T, s}$ = 10 s, as obtained from the Stokes--Einstein equation for translational diffusion D$_{T, s}$ = k$_B$T / 6$\pi\eta R_s$ assuming a particle of the same size as the satellite orbiting a circular path, with $\rho$ the orbit radius from the measured satellite-core~distance.\\

{\paragraph*{Modeling}--}
To verify the leading mechanisms that couple the self-reconfiguration of a molecule to its reorientation, we describe satellite and core particles with the same size ratio as in the experiments by two-dimensional diffusive objects with dynamics satisfying the Einstein relations. In brief, when a satellite, by diffusion, approaches a core particle closer than a cutoff distance, strong attractive interactions lead to the formation of a molecule which breaks translational symmetry causing active motion. The asymmetric flow around the molecule, modelled by an effective force, induces an active velocity with direction along the line joining the centers of the satellite and core. With these minimal ingredients, the satellite can diffuse around the core and its angular position determines the direction of the propulsion velocity of the whole molecule, as in the experiments. As a result, the present model goes beyond standard active Brownian particle dynamics, where particle orientation would only be governed by the rotational diffusion of the whole molecule as a rigid body. Orientation is here intrinsically controlled by the instantaneous configuration of the molecule driven by satellite diffusivity, see Appendix for details on the model. 

Our model quantitatively reproduces the timescale for reorientation measured experimentally in both the self-reconfiguring and pre-assembled cases (Fig.~\ref{fig:fig2}c), and predicts a systematic dependence of $\tau_R$ of the satellite's diffusivity beyond the experimental realizations (Supplemental Fig. S3). In particular, we verify the hypothesis that $\tau_R$ decreases as the satellite translational diffusion $D_{T, s}$ increases, i.e. that the timescale for reorientation is mainly determined by the diffusive motion of the satellite, which induces the reconfiguration of the molecule. Since $D_{T, s}$ can be controlled via tuning the satellite size according to the Einstein relation, this finding is highly relevant for designing complex molecules with controllable reorientation dynamics.\\

\begin{figure*}[hb!]
    \centering
     \includegraphics[width=1\linewidth]{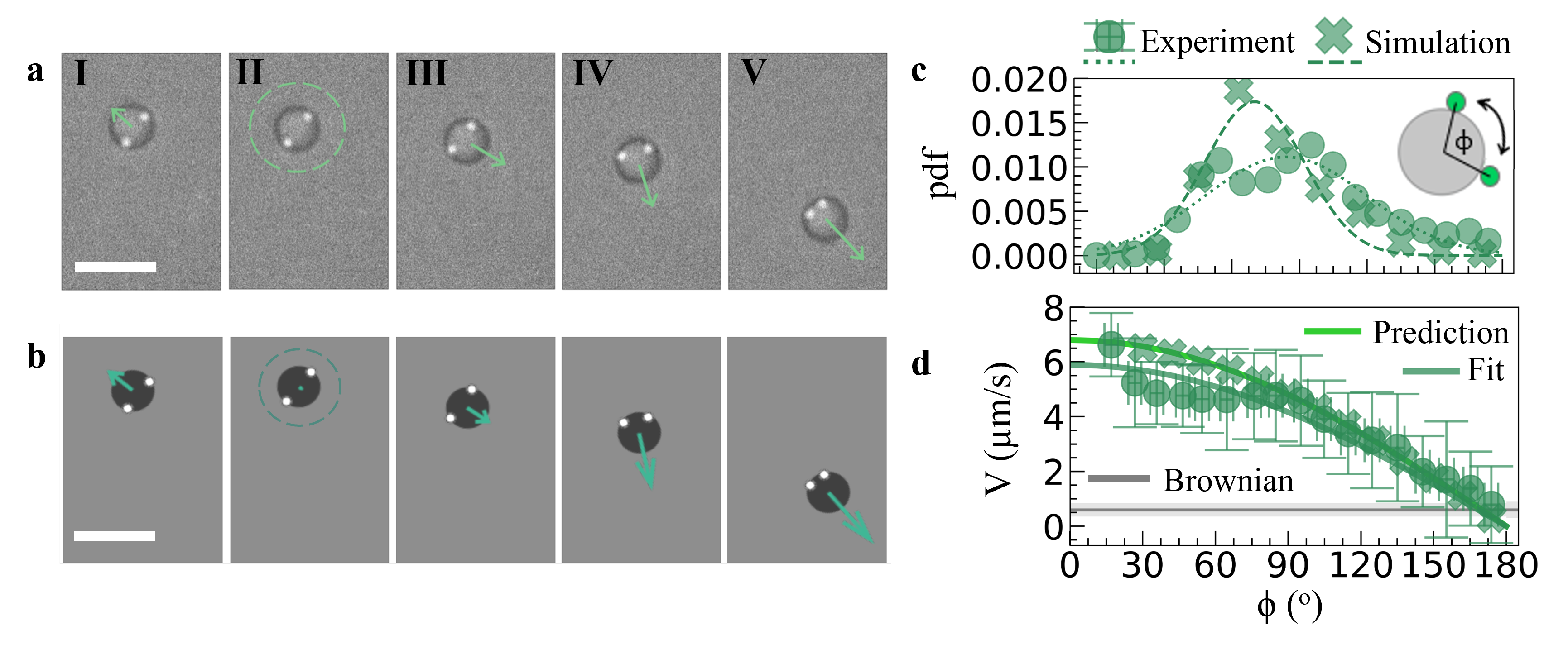}
    \vspace{-25pt}
    \caption{\textbf{Reconfiguration-induced reorientation at the single-molecule level.} \textbf{a-b,} Time series of a reconfiguring active trimer in \textbf{a,} the experiments and \textbf{b,} the simulations. Arrows indicate the magnitude of the velocity. Over time, satellites probe different configurations, which result in directed motion -- when they are distributed asymmetrically with respect to the core -- or a stationary state -- for symmetric configuration -- followed by reversal of the direction of motion when both satellites cross over to one side of the core. \textbf{c,} Probability density function of the opening angle $\phi$ between satellites. Dotted and dashed lines are fitted Gaussian distributions. \textbf{d,} Instantaneous velocity of reconfiguring trimers as a function of the instantaneous opening angle between the satellites in experiments and simulations. The velocity decreases with opening angle following a simple geometric argument as presented in the text. Errors are calculated from standard deviations as described in the text.}
    \label{fig:fig3}
\end{figure*}

{\paragraph*{Molecules with complex composition}--} To unravel further implications of reconfiguration on active motion, we then examine complex colloidal molecules with two or more satellite particles as in Movie 2. Figure~\ref{fig:fig3}a shows a measured time series of a self-reconfiguring active trimer with two satellites. We observe that, as satellites diffuse around the core, they continuously probe different instantaneous configurations. 

Satellites can be distributed either asymmetrically relative to the core (column I) or symmetrically around it (column II). The asymmetric configuration leads to symmetry breaking of the EHD flows, and correspondingly to self-propulsion with the satellites at the back, as for molecules with a single satellite. The symmetric configuration leads to a radially symmetric fluid flow and an instantaneous stationary state. After going through a symmetric configuration, a change in the net direction of motion can be observed (column III) and the molecule self-propels in a new direction defined by the subsequent asymmetric configurations (columns IV and V). We reproduce the motion of reconfiguring trimers in our simulations imposing that each satellite provides an active force and contributes to determining the total molecule dynamics. In Fig.~\ref{fig:fig3}b, we recover the experimentally observed configurational changes by introducing an effective long-range repulsion between satellites, effectively induced by hydrodynamics interactions.

Qualitatively, the instantaneous configuration of the satellites affects self-propulsion both in the experiments and simulations. We subsequently quantify the continuously fluctuating opening angle \emph{$\phi$} between satellites in the experiments and correlate it to the magnitude of the active velocity. The probability density function of opening angles obtained from different trimers with two satellites shows a bell-shaped distribution which resembles a Gaussian (Fig.~\ref{fig:fig3}c), commensurate with the Brownian nature of the satellites' motion, with a peak value $\phi_{P,~Exp}$ = (84 $\pm$ 36)$^o$ determined by a least-squares fit (dotted line). The experimental behavior is well captured by our simulations, with $\phi_{P,~Sim}$ = (70 $\pm$ 22)$^o$. Once the instantaneous configuration of the molecules is known, the corresponding instantaneous active velocity is determined. More than 8000 data points from different trimers are grouped together and binned by opening angle (bin width = 10$^o$, field frequency 1 kHz, amplitude 6 $V_{pp}$). We fit the velocity and angle distributions within each bin with a Gaussian and report the fitted mean and standard deviation in Fig.~\ref{fig:fig3}d. Assuming that each satellite contributes equally to the active force by the corresponding EHD flows, we predict that the velocity will follow the ``sum rule'' expression $V_{pred}$ = 2$V_1$cos($\phi$/2), with $V_1$ corresponding to the experimental velocity of self-reconfiguring dimers at the same field condition ($V_1$ = (3.4 $\pm$ 0.7) \textmu m/s, error denotes standard error). We show that this geometrical sum accurately describes our measurements in Fig.~\ref{fig:fig3}d (green line ``Prediction''). Fitting the data while keeping $V_1$ as an open fit parameter yields V$_{fit}$= (3.0 $\pm$ 0.1) \textmu m/s (green line ``Fit'', error obtained from the covariance matrix), in agreement with the measured $V_1$. Simulations reproduce this behavior in Fig.~\ref{fig:fig3}d. We confirm that the sum rule of the velocity also holds for molecules with different numbers of satellites, e.g. see Supplemental Fig. S4 for data on tetramers with three satellites. At full valency, i.e. 11 units here (endecamer), no net directed motion is observed, as expected by symmetry. Consequently, manipulation of both the composition and configuration of the molecule would allow tailoring the active velocity. 

\subsection*{\large Reconfiguration-induced self-steering and avoidance}

\begin{figure*}[hb!]
    \centering
     \includegraphics[width=1\linewidth]{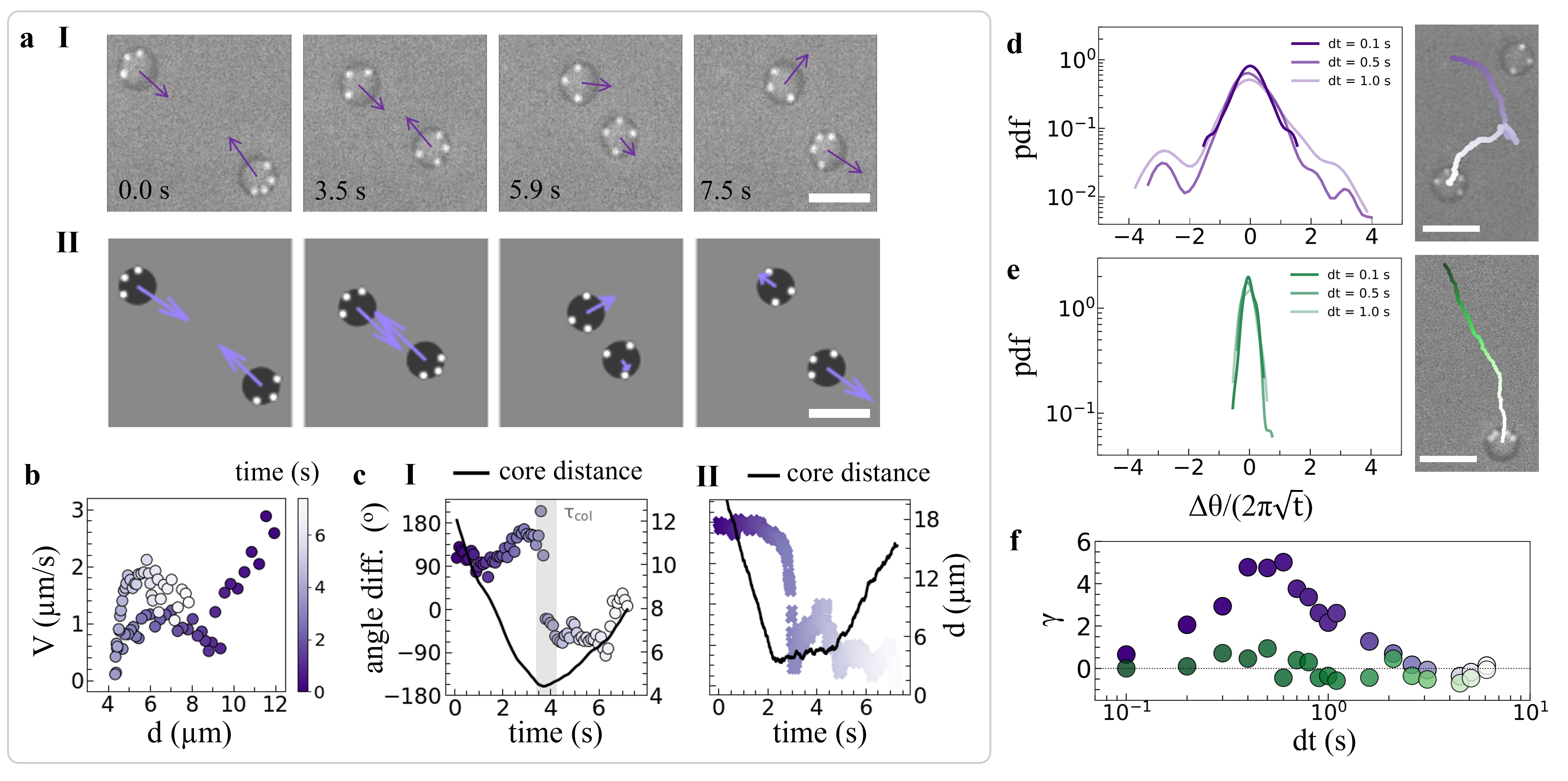}
    \vspace{-20pt}
    \caption{\textbf{Reconfiguration-induced avoidance.} \textbf{a,} Time series of interacting molecules from I) the experiments and II) simulations. Arrows indicate the direction and magnitude of instantaneous velocities. \textbf{b,} Velocity of a self-reconfiguring pentamer with four satellites as a function of distance from its neighbor. Time is color-coded according to the vertical bar. \textbf{c,} Angle difference between the orientation of the velocity of interacting molecules as a function of time measured in I) experiments and II) simulations. Secondary $\psi$-axis shows the intermolecular distance; the shaded region highlights the duration of the ``collision''. Distribution of angular displacements of the core of a self-reconfiguring pentamer at different lag-times and corresponding active trajectory \textbf{d,} in the presence of interactions, and \textbf{e,} without interactions. \textbf{f,} Kurtosis of the angular displacement distributions at different lag-times with and without interactions. Scale bars are 10~\textmu m.}
    \label{fig:fig4}
\end{figure*}

We next turn to the implications of self-reconfiguration at the level of the interactions between microswimmers, seeking to understand the role of reconfiguration-induced reorientation on self-organization at the collective level. Strikingly, when two of our active molecules approach, they sense each other's presence at a distance and respond by changing propulsion direction through spontaneous reconfiguration (Movie 3). This is exemplified in Fig.~\ref{fig:fig4}a-I.

Initially, the molecules self-propel toward each other with the satellites at the back. However, upon coming within a critical separation distance, the respective satellites experience a long-ranged attraction to the neighboring core and correspondingly reconfigure, seemingly acting as sensors. With decreasing distance, repulsion between the cores takes over and prevents contact between molecules or satellite exchange. Nonetheless, at this stage, the satellites have migrated to the other side of the core, thus inducing a reversal of propulsion direction, leading the particles to self-steer away from each other. This change in direction happens much faster ($\lesssim$ 1 s) than the previously discussed reconfiguration-driven reorientation without interactions, while it is still mediated by satellite reconfiguration. In essence, interacting molecules actively exhibit reconfiguration-induced avoidance, reminiscent of self-avoidance observed in motile organisms across length scales and different from avoidance due to pure repulsion as we discuss below.

We quantify the different stages of self-avoiding interactions by measuring the velocity of one of the two molecules (the pentamer in Fig.~\ref{fig:fig4}a-I) in Fig.~\ref{fig:fig4}b, and the angle between the velocity vectors of the two molecules as a function of inter-molecule distance in Fig.~\ref{fig:fig4}c-I. Upon approach, at large separation distance (color-coded in the side bar as a function of elapsed time), the velocity decreases as the molecule's satellites move toward the other side of the core via configurations that are more symmetric with respect to the center, which correspondingly slow down the molecule in line with the results of Fig.~\ref{fig:fig3}. Repulsion between cores, mediated by electro-hydrodynamic flows~\cite{Yang2017}, additionally contributes to hindering the forward motion of the molecule. The velocity goes to zero at a minimum distance ($\approx$ 4.5 \textmu m), at which the satellites become tightly packed at the opposite hemisphere, pushing the molecules in the opposite direction. As the distance increases, satellites cease sensing the presence of a neighboring particle and resume Brownian diffusion along their core. The reversal in the sign of the velocity vectors at a minimum distance in Fig.~\ref{fig:fig4}c-I clearly indicates anti-alignment (note that, for convention, we define positive and negative angles as the aligned and anti-aligned states, respectively). 

However, since the direction reversal happens faster than natural reorientation, these anti-aligning interactions affect the distribution of angular displacements in the propulsion direction differently at different lag-times dt (Fig.~\ref{fig:fig4}d). In particular, at time scales shorter than the particle reconfiguration, e.g. dt = 0.1 s, the angular displacements follow a Gaussian distribution centered around zero, as for active Brownian particles, whose width is defined by the satellites' D$_{T,s}$ as initially discussed. On the other hand, at lag-times comparable to the characteristic duration of ``collisions'', large angular-displacement tails emerge (dt = 0.5 s and dt = 1.0 s). 

As a comparison, the same analysis on a self-reconfiguring pentamer that does not interact with any other molecule along its active trajectory shows no sign of non-Gaussian behavior (Fig.~\ref{fig:fig4}e). The evolution of non-Gaussian orientational dynamics at different lag-times is readily quantified by the kurtosis $\gamma$ of the distributions of angular displacements, plotted in Fig.~\ref{fig:fig4}f; $\gamma$(dt)$>$0 corresponds to heavy tails and $\gamma$(dt)=0 to a Gaussian distribution. We observe that the kurtosis peaks at 0.6 s, comparable to the collision time, and that this behavior persists across different molecules with varying compositions undergoing multiple collisions (Supplemental Fig. S4). We furthermore fully reproduce this behavior in simulations (see Fig.~\ref{fig:fig4}a-II and~\ref{fig:fig4}c-II and Appendix for details on modeling interactions) confirming that the interplay between the molecule reconfiguration and satellites-core interactions are responsible for effective anti-alignment interactions. This confirmation additionally corroborates the mechanism underpinning reconfiguration-induced avoidance in our system and in turn determines minimal ingredients for its realization.

\begin{figure*}[ht!]
    \centering
     \includegraphics[width=1\linewidth]{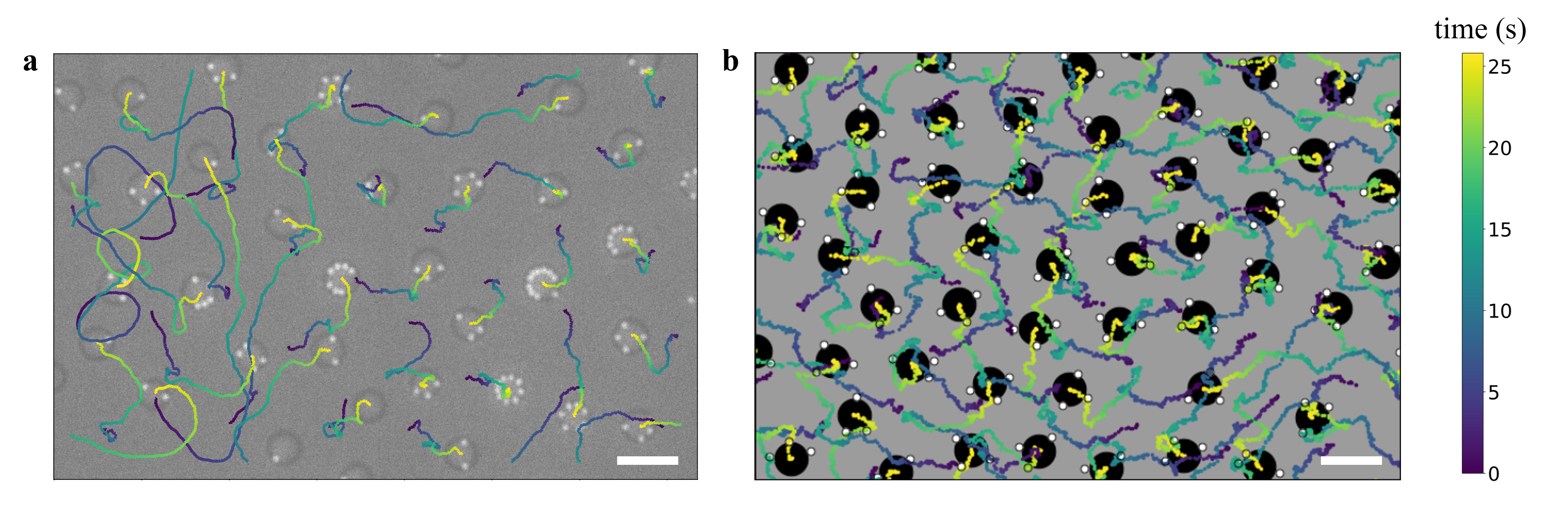}
    \vspace{-20pt}
    \caption{\textbf{Collective dynamics: reconfiguration-induced avoidance hinders activity-induced clustering, phase separation and flocking at high area fractions.} Self-reconfiguring molecules uniformly occupy space while allowing for collective rearrangements without activity-induced clustering in both \textbf{a,} experiments, with molecules of varying compositions, and \textbf{b,} simulations of tetramers with three satellites. Area fractions are 0.11 and 0.16, respectively. All lines represent trajectories color-coded with time. Scale bars are 10 \textmu m.}
    \label{fig:fig5}
\end{figure*}

\subsection*{\large Discussion and outlook}\label{discussion}

We present an experimental system of active colloidal molecules that exhibits autonomous steering and avoidance through self-reconfiguration. Internal reconfiguration decouples reorientation from rotational properties and sets their instantaneous velocity, such that translational and rotational motions are effectively independent. Thus, these molecules are governed by distinctly different dynamics compared to traditional active colloids. Active reconfiguration here leads to speed modulation and reorientation on the single-molecule level as well as flexible organization on the collective level.

Standard active colloids with a constant velocity and fixed rotational diffusivity phase separate or form dynamic clusters at area fractions as low as $\approx$~3\%~\cite{Theurkauff2012,Buttinoni2013,Ginot2018}, while our reconfiguring molecules remain well-separated up to the highest investigated area fractions of~$\approx$~16\% (examples in Fig.~\ref{fig:fig5} and Movie 4). Intriguingly, at the collective level the qualitative dynamics of the system does not appear to be strongly affected by the number of satellites of the active molecules, as observed by comparing the experiments to the monodisperse case of active tetramers with three satellite particles from the simulations (Fig.~\ref{fig:fig5}b). 

Reconfiguration-induced avoidance here promotes a uniform spatial distribution of the particles while maintaining fluidity and allowing for collective rearrangements driven by activity. This active type of avoidance differs from avoidance due to repulsion alone observed recently in rigid active colloids which led to flocking and formation of active Wigner crystals~\cite{Das2024}. Our system thus constitutes a promising candidate for studying collective dynamics in active glassy systems~\cite{Janssen2019}, or for constructing hyperuniform systems~\cite{Huang2021, Boltz2024}. In particular, reconfiguration-induced avoidance tends to homogenize particle distribution such that density fluctuations are strongly suppressed, which is a prerequisite for hyperuniformity~\cite{Lei2019}. 

Finally, the type of anti-alignment after collisions that we report here is in contrast with the requirements to establish the classical, previously-observed motility-induced phase separation, as well as typical flocking. In those cases, alignment promotes the transition, even in the case of repulsive interactions~\cite{Das2024}. Anti-aligning interactions can conversely induce completely different patterns, as recently explored by theory~\cite{Escaff2024feb, Escaff2024aug, Arold2020}, including dancing hexagons and exotic wavy patterns. Reconfiguration-induced anti-alignment has insofar remained unexplored, noting that in previous studies~\cite{Zion2023, Mirhosseini2022} single particles experienced a torque, which changed their orientation relative to the propulsion direction.

We therefore propose that active reconfiguration can prove useful in future attempts to realize particles with advanced navigation abilities inside complex environments, a key step towards biologically-inspired functional synthetic systems. Autonomous steering and avoidance through reconfiguration can open the door towards novel modes of transport and active response to surfaces, beyond particle guiding by obstacles and self-trapping at topographic boundaries observed in current synthetic active matter realizations, where particle escape is dictated by the topography of the environment~\cite{Ketzetzi2022}.

\subsection*{\large Acknowledgements}
LI gratefully acknowledges funding from the European Research Council (ERC) under the European Union's Horizon 2020 research and innovation program grant agreement No.~101001514. HL acknowledges funding from the German Research Foundation (DFG) within LO 418/29-1. SK thanks Vincent Niggel for SEM imaging, Steven van Kesteren for inspiring the pre-assembled molecules' fabrication procedure, Federico Paratore for electrode coating in preliminary experiments, Robert Style and Ueli T\"{o}pfer for assistance with code. 

\clearpage

\subsection*{\large Appendix I: Methods}
\subsection{Colloidal molecules} 
Polystyrene particles with diameter 0.71 $\pm$ 0.02 {\textmu}m (green fluorescent) and silica particles with diameter 5.64~$\pm$~0.27~{\textmu}m were purchased from Microparticles GmbH. Pre-assembled molecules were formed by mixing SiO$_2$ and PS particles in MilliQ water containing 50 mM HCl for 15 min while shaking at 500 rpm at 80$^o$ (SiO$_2O$ 0.1 wt \%, PS 0.001 wt \%), followed by thermal sintering in the oven for 15 minutes at 90$^o$ and harvesting with a water droplet into the experimental cell. SEM imaging showed that this process overall did not affect the particles. Self-reconfiguring molecules were formed inside the experimental cell described below in the following manner: SiO$_2$ ($\approx$ 0.005 wt\%) and PS ($\approx$ 0.0005 wt\%) particles were mixed in MilliQ water and placed in the cell, where they sedimented above the bottom electrode. Self-assembly was triggered under the AC electric field for $V_{pp}$$>$2V at a frequency of 1 kHz. The initial particle concentration is not crucial as variations in molecule composition are naturally observed throughout the sample, albeit increasing the PS concentration triggers formation of molecules with full valency that do not display directed motion (cores with $\simeq$1 ring of PS).

\subsection{Experimental setup and imaging} 
The particle suspension was placed in a custom-made sample cell that consisted of two transparent electrodes separated by an adhesive spacer with a 9 mm-circular opening and 120 \textmu m height (Grace Bio-Labs SecureSeal). The electrodes were glass slides (borosilicate glass 18x18 mm No2 Corning) coated via electron beam metal evaporation with 3 nm Cr and 10 nm Au (Plassys MEB550S) and plasma enhanced chemical vapor deposition with 10 nm of SiO$_2$ (Multiplex CVD, Oxford PlasmaPro 100) to reduce particle adhesion on the conductive glass slide. The electrodes were connected to a function generator (National Instruments Agilent 3352X) that applies the AC electric field (frequency 1 kHz, $V_{pp}$ 2-6V). Particles were imaged using x20, x40 and 63x air objectives (Zeiss) mounted on an inverted microscope (Axio Observer D1) with an sCMOS camera (Andor Zyla) at a frame rate of 10 fps under bright-field transmission illumination in combination with epifluorescence illumination (North 89, filter AF488 nm -- Excitation: 450 - 490 nm, Emission: 515 - 1100 nm) to simultaneously identify both SiO$_2$ and fluorescently-labelled PS particles comprising the molecules. Image sequences were pre-processed on ImageJ and subsequently analyzed using standard particle tracking algorithms in Python or Matlab.

\setcounter{figure}{0}
\newcommand{\bluec}[1]{\textcolor{blue}{[#1]}}
\newcommand{\redc}[1]{\textcolor{red}{[#1]}}
\renewcommand{\thefigure}{S\arabic{figure}}

\subsection{Calculation of electrohydrodynamic flows and molecule velocity}

As discussed in the main text, our colloidal molecules propel themselves relative to the electrode due to the asymmetry in the electrohydrodynamic (EHD) flow around the SiO$_2$ particles caused by the PS particles' self-assembly. The velocity of a dimer molecule, $\rm V_1$, comprising one SiO$_2$ core particle and one PS satellite particle can be obtained in first order approximation as a linear combination of the EHD flow velocities $\rm U_i$ generated around each particle “i”, evaluated as a function of distance, r, away from each particle surface \cite{Ma2015,Ma2015PNAS}:

\begin{equation}\label{eq:eq1}
    U_i = \frac{C~(K_i ^{'} + K_i ^{''}) 
 ~\bar{\omega} }{\eta ~(1 +~\bar{\omega}^2)} \frac{3~(r/R_i)}{2~[1+(r+R_i)^2]^{5/2}}
\end{equation}

\begin{equation}\label{eq:eq2}
    V_1 = \frac{U_{SiO_2}~R_{PS} + U_{PS}~R_{SiO_2}}{R_{PS} + R_{SiO_2}}
\end{equation}

\noindent with $\eta$ the medium viscosity, R$_i$ the particle radius, r = R$_{PS}$ + R$_{SiO_2}$, $ C=\beta \epsilon_m \epsilon_0 H (V_{pp}/2H)^2 $, $\beta$ a constant prefactor used here as a single fit parameter, $\epsilon_m$ the solvent relative permittivity, $\epsilon_0$ the vacuum permittivity, $\bar{\omega}$ the normalized angular frequency $\bar{\omega} = \omega H/\kappa D$ with $\omega = 2 \pi f$ and f the frequency of the applied electric field, $\kappa$ the inverse Debye length and D the ion diffusivity. 

\begin{figure*}[hb!]
    \centering
    \includegraphics[width=0.8\linewidth]{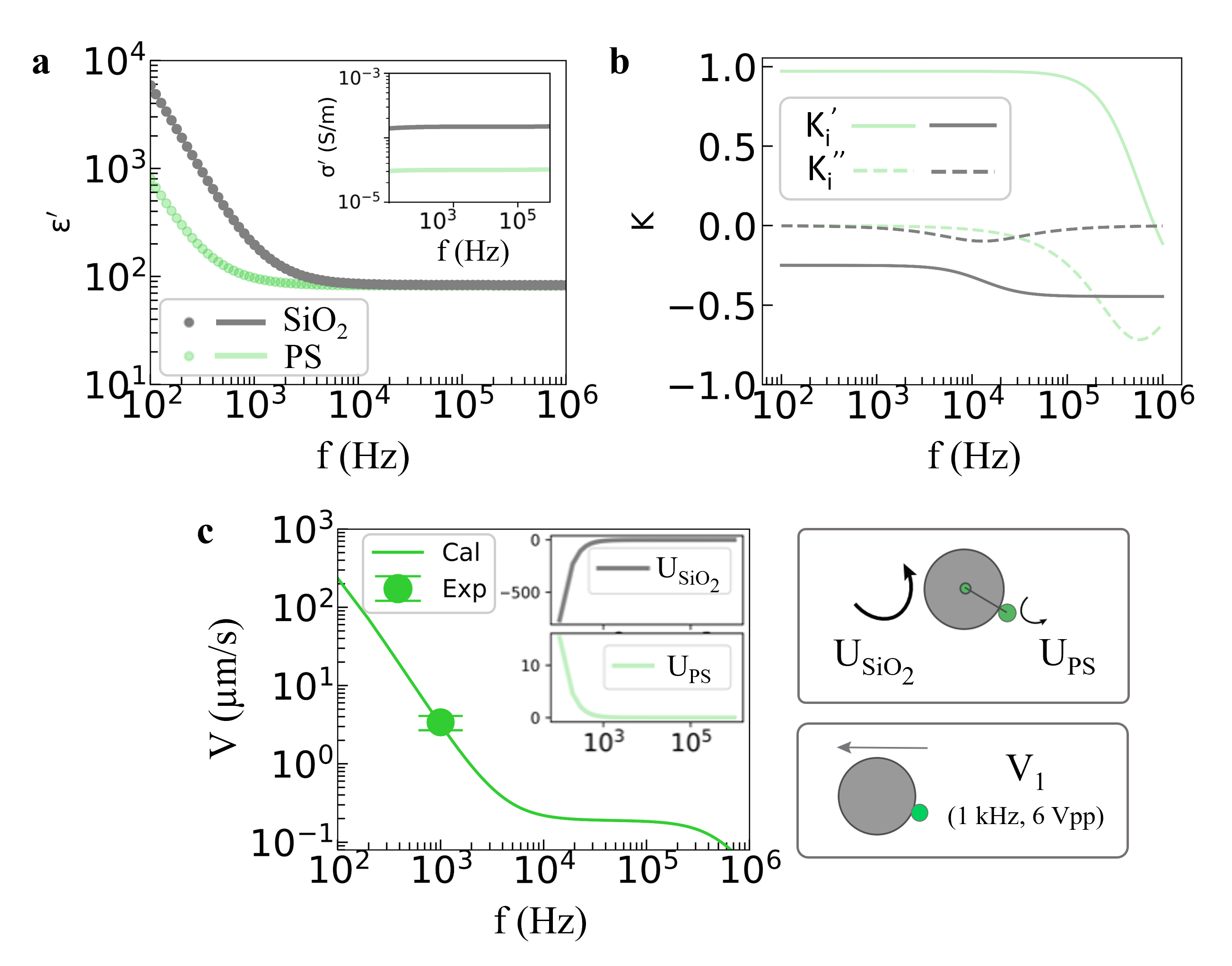}
    \caption{\textbf{Dielectric properties and polarizability of the SiO$_2$ and PS colloids under study.} \textbf{a,} Measured polarizability $\epsilon'_p$ (main) and surface conductivity $\sigma'_p $ (inset) a function of frequency for each particle type at temperature 24$^o$. \textbf{b,} Calculated Clausius-Mossotti (complex polarizability) factor as a function of frequency for each particle type. The values are obtained using the corresponding values in \textbf{a}. \textbf{c,} Main panel: net molecule velocity V$_1$ as a function of frequency using Eq.~\ref{eq:eq2}, with the single fit parameters $\beta_{PS}$ and $\beta_{SiO_2}$ 1.3 and 1.6, respectively. Inset: theoretically predicted EHD flow velocities for each particle type, negative for the case of SiO$_2$ and positive for the PS, using the corresponding values in \textbf{b} and Eq.~\ref{eq:eq1}. Schematics indicate the direction of the positive (i.e., repulsive in the case of PS) and negative (i.e., attractive in the case of SiO$_2$) flows.}
    \label{fig:figS1}
\end{figure*}

$K'$ and $K''$ are the real and imaginary part of the Clausius-Mosotti factor $K^{\ast} = K' + jK''$ \cite{Pethig2017} with $j=\sqrt{-1}$. They play a key role in $U_{\rm i}$ as they determine the sign of the EHD flows, as well as contributing to their magnitude. The Clausius-Mossotti factor in fact describes the polarizability of a particle suspended in a fluid and it dictates the magnitude and sign of the induced dipole moment and therefore the distortion of the fluid flow of a charged particle under an AC electric field \cite{Shilov2000,Pethig2017}. In particular:
\begin{eqnarray}
    K' &=& \frac{\omega^2 \epsilon_{\rm 0}^2(\epsilon'_{\rm p} - \epsilon'_{\rm m}) (\epsilon'_{\rm p} + 2\epsilon'_{\rm m}) + (\sigma'_{\rm p} - \sigma'_{\rm m}) (\sigma'_{\rm p} + 2\sigma'_{\rm m})}{\omega^2 \epsilon_{\rm 0}^2(\epsilon'_{\rm p} + 2 \epsilon'_{\rm m})^2 + 2(\sigma'_{\rm p} + 2 \sigma'_{\rm m})^2}
    \label{K1}
\\[2mm]
    K'' &=& \frac{\omega \epsilon_{\rm 0}(\epsilon'_{\rm p} - \epsilon'_{\rm m}) (\sigma'_{\rm p} + 2\sigma'_{\rm m}) - \epsilon_{\rm 0}(\epsilon'_{\rm p} + 2\epsilon'_{\rm m}) (\sigma'_{\rm p} - \sigma'_{\rm m})}{\omega^2 \epsilon_{\rm 0}^2(\epsilon'_{\rm p} + 2 \epsilon'_{\rm m})^2 + 2(\sigma'_{\rm p} + 2 \sigma'_{\rm m})^2}
    \label{K2}
\end{eqnarray}

\noindent with $\epsilon_p^{'}$ and $\epsilon_p^{'}$ the real part of relative particle and medium permittivity, $\sigma_p^{'}$ and  $\sigma_m^{'}$ the real part of particle and medium conductivity. We show the $\epsilon'_p$ and $\sigma'_p$ for both particle types in Fig.~\ref{fig:figS1}a. These values were obtained from renormalization of dielectric spectroscopy measurements of the effective $\epsilon_{eff}^{'}$ and $\sigma_{eff}^{'}$ values of each individual particle suspension as a function of frequency (10 to 10$^6$ Hz) at temperature 24$^o$ (Novocontrol high-resolution dielectric analyzer Alpha-A). For this, we considered Maxwell-Wagner-Sillars' theory \cite{Alvarez2021}, where the complex dielectric function of a heterogeneous mixture of spherical particles relates to the dielectric properties of its components as:
\begin{equation}
    (\epsilon_\textrm{eff}^{\ast} - \epsilon_m ^{\ast}) / (\epsilon_\textrm{eff}^{\ast} + 2\epsilon_m^{\ast}) = \phi  (\epsilon_p^{\ast} - \epsilon_m^{\ast})/ (\epsilon_p^{\ast} + 2\epsilon_m^{\ast}) = \phi ~ K^{\ast},
\end{equation}
with $\phi$ the volume fraction of the particles dispersed in the matrix medium (1 wt \% for PS particles and 0.1 wt \% for SiO$_2$ particles), $K^{\ast}$ the Clausius-Mossotti factor of the particles in the suspension, and $\epsilon_\textrm{eff}^\ast$, $\epsilon^{\ast}_{p}$, $\epsilon^{\ast}_{m}$ the complex permittivity of the suspension, the particle and the medium, respectively.

Using the measured $\epsilon'_p$ and $\sigma'_p$ values of Fig.~\ref{fig:figS1} we then calculate $K'_i$ and $K''_i$. The behavior of K in Fig.~\ref{fig:figS1}b already suggests that for low f values, i.e., 1 to 10 kHz regime, interactions via EHD flows dominate and dipolar interactions are negligible. This is due to strong ionic screening effects, as the surface ions follow the polarization timescale of the applied electric field. As f increases moderately to $>$ 20 kHz, ionic screening is largely reduced, such that there is a competition between EHD and dipole interactions. In the MHz regime, ions cannot follow the rapidly oscillating field, and the bulk material response dominates, with particle interactions described by dipole-dipole interactions. 

Thus, the main interaction dominating in experiments performed at 1 kHz is based on the attractive EHD flows around the SiO$_2$ particles, bringing the PS towards it and causing their self-assembly by geometrically yet flexibly trapping the PS between the electrode and the equatorial plane of the SiO$_2$ particle. 

Using the $K'_i$ and $K''_i$ values of Fig.~\ref{fig:figS1}b, we predict $U_i$ for each particle of the dimer evaluated at a position r corresponding to the center of the other particle using Eq.~\ref{eq:eq1}. The calculated relative flow velocities, U$_{SiO_2}$ and U$_{PS}$, are shown as insets in Fig.~\ref{fig:figS1}c together with a schematic representation (not to scale). The calculation shows that the magnitude of the U$_{SiO_2}$ flow velocity decreases with increasing frequency. The flow is pointing inwards in the direction perpendicular to the electrode, as evidenced by the negative sign. Note that the magnitude of U$_{PS}$ is considerably lower. Finally, we combine the flow velocities for each particle to obtain the dimer velocity V$_1$ in the main panel of Fig.~\ref{fig:figS1}c. This is obtained as a linear combination, see Eq.~\ref{eq:eq2}, and matches the experimentally measured instantaneous velocity of self-reconfiguring dimers in the manuscript.\\

\subsection{Model for self-reconfiguring active molecules}

An active self-reconfiguring molecule is a many-particle colloidal object in solution consisting of a core, i.e.\ a big colloidal particle with radius $R_c$, and $N$ satellites, i.e.\ small colloidal particles with radius $R_s$. Satellites assemble around the core and are free to orbit along its longitudinal axis. Thus, once a molecule with $N$ satellites is formed, asymmetric configurations in satellite arrangement can be observed. This molecule is considered active since it shows a swimming motion along the asymmetric axis determined by the relative position of satellites compared to the core.

Prior to molecule formation, core and satellite particles behave as passive objects, being symmetric colloids immersed in a solution that satisfies the Einstein relation. As such, they can be described by overdamped dynamics for their position, governed by the diffusion coefficient $D$ given by:
\begin{equation}
D_{\alpha}= \frac{k_B T}{6\pi \eta R_{\alpha}} \,,
\end{equation}
with $k_B$ the Boltzmann constant, $T$ the room temperature, and $\eta$ the water dynamic viscosity. The term $R_\alpha$ represents the particle radius, with $\alpha=c, s$ for core and satellite particles, respectively.

\vskip0.2cm
\noindent
{\bf Molecule dynamics.} 
Since experiments show quasi-two-dimensional motion for the active molecules, here, we neglect the third dimension and restrict the dynamics to two dimensions. The dynamics of the system is determined by overdamped equations of motion for core and satellites mutually interacting through a potential. The experimentally observed active motion is generated by a non-conservative force that cannot be obtained from a potential. Specifically, the dynamics for the core particle position $\mathbf{x}_{c_j}$ and satellite particle positions $\mathbf{x}_i$ read
\begin{subequations}
\label{eq:molecule_dynamics}
\begin{align}
& \dot{\mathbf{x}}_{c_j}= \sqrt{2D_c} \boldsymbol{\xi}_{c_j} + \frac{\mathbf{F}_{c_j}}{\gamma} + V_1 \mathbf{n}_j  \\
&  \dot{\mathbf{x}}_i= \sqrt{2D_s} \boldsymbol{\xi}_i + \frac{\mathbf{F}_i}{\gamma}+  \frac{\mathbf{F}^r_i}{\gamma} +  V_1 \mathbf{n}_j \,,
\end{align}
\end{subequations}
where the index $i=1, ..., N$ identifies the $i$-th satellite of an active molecule consisting of a core and $N$ satellites, while the index $j=1, ..., n$ denotes the $j$-th core particle and therefore $j$-th molecule. In general, we use the pedex $c$ to denote variables and forces characterizing the core particle.
The terms $\boldsymbol{\xi}_{c_j}$ and $\boldsymbol{\xi}_i$ are independent white noises with zero average and unit variance.
In this dynamics, the constant $\gamma$ represents the effective friction coefficient due to the water solution so that $\mathbf{F}_{c_j}$ and $\mathbf{F}_i$ are the forces acting on core and satellites, respectively. Finally, $V_1 \mathbf{n}_j$ corresponds to the active velocity of the molecule and, thus, governs the dynamics of both cores and satellites.

\vskip0.2cm
\noindent
{\bf Intra-molecule interactions. } 
As experimentally shown, when satellite and core particles are close they attract each other and form a stable molecule.
This mechanism is modeled by considering an attracting force $\mathbf{F}_i$ between the satellite $i$ and the core particle $j$.
Consequently, the force $\mathbf{F}_{c_j}$ acting on the core is due to the contribution of all the molecule satellites and is given by $\mathbf{F}_{c_j}= - \sum_{i=1}^{i=N} \mathbf{F}_i$. This attraction is due to a Lennard-Jones potential $\mathbf{F}_i=-\nabla_i U_{LJ}(|\mathbf{x}_i -\mathbf{x}_c|)$, which constrains the satellite center of mass position to be at distance $\sigma_{cs}=R_c + R_s$ from the center of mass position of the core particle. Explicitly, $U_{LJ}(|\mathbf{x}_i -\mathbf{x}_c|)$ can be expressed as:
\begin{equation}
U_{LJ}(r)=
4\epsilon_{cs} \left[\left(\dfrac{\sigma_{cs}}{r}\right)^{12}-\left(\dfrac{\sigma_{cs}}{r}\right)^{6}\right], \quad r<3\sigma_{cs}
\end{equation}
and zero otherwise for $r\geq3\sigma_{cs}$.
This potential is cut at distance $r=3\sigma_{cs}$ and additionally shifted by the constant $-U_{LJ}(3\sigma_{cs})$ so that it is continuous in zero.
The term $\epsilon_{cs}$ determines the typical energy scale of the potential which also fixes the potential barrier to observe the molecule breaking.
Therefore, the molecule stability is guaranteed if thermal fluctuations are smaller compared to the potential barrier due to the attractive force $\mathbf{F}_{c_j}/\gamma$, with $\gamma$ the effective friction coefficient of water. 
This implies that the molecule is stable if the following condition holds $D_s\ll \epsilon_{cs}/\gamma$.

Additionally, satellites repel each other with an extra force $F^r_i=-\nabla_i U^r_{tot}$ due to a pure-repulsive potential $U^r_{tot}=\sum_{i<j} U^r(|\mathbf{x}_i -\mathbf{x}_j|)$. 
The shape of $U^r(r)=U^d(r)+U_{w}(r)$ consists of a dipole-dipole repulsive potential and a Weeks-Chandler-Anderson repulsive potential. The first is defined as
\begin{equation}
U^d(r)=
\epsilon_{d}  \left(\dfrac{\sigma_{s}}{r}\right)^{3}, \quad r<5\sigma_{s}
\end{equation}
and reads zero for $r>5\sigma_{s}$, while the second is defined as
\begin{equation}
U_{w}(r)=
4\epsilon_{w} \left[\left(\dfrac{\sigma_{s}}{r}\right)^{12}-\left(\dfrac{\sigma_{s}}{r}\right)^{6}\right], \quad r<2^{1/6}\sigma_{s}
\end{equation}
and zero otherwise.
In addition, $\sigma_s=2R_s$ is the satellite diameter, while $\epsilon_d$ and $\epsilon_{w}$ represent the typical energy scale of the dipole-dipole and WCA repulsive potentials, respectively.
Finally, both potentials are shifted by irrelevant constants, $C^d=-U^d(5\sigma_{s})$ and $C_w=-U_{w}(2^{1/6}\sigma_{s})$, respectively, so that both are continuous functions.

\vskip0.2cm
\noindent
{\bf Active velocity.}
The term $V_1\mathbf{n_j}$ models the experimentally observed effective driving velocity of the j-th molecule, which affects the evolution of the core and satellites.
Specifically, the constant $V_1$ is the velocity scale observed in the simpler (dimer) molecule, consisting of a core and one satellite, and
is chosen so that the measured dimer molecule velocity is reproduced by numerical simulations.
The vector $\mathbf{n_j}$ identifies the direction of the active velocity. This is not a unit vector as it additively accounts for the contribution of each satellite.
It is defined as
$\mathbf{n}_j=\sum_{i=1}^N\mathbf{n}_{ij}$
with $\mathbf{n}_{ij}$ given by:
\begin{equation}
\mathbf{n}_{ij}= \theta(r_p - r_{ij}) \frac{\mathbf{r}_{ij}}{r_{ij}^3} \sigma_{cs}^2 \,.
\end{equation}
Here, $\mathbf{r}_{ij}=\mathbf{x}_i -\mathbf{x}_{c_j}$ is the vector pointing between the core $j$ and the satellite $i$ while $r_{ij}=|\mathbf{r}_{ij}|$ is its modulus.
The function $\theta$ selects satellites whose distance is smaller than $r_p$ from the core.
Each satellite contributes to the non-conservative force proportional to the effective velocity $V_1$, the typical swim velocity of the molecule, which is additionally modulated in space by a function of the distance.
The choice of the force shape scaling as $1/r_{ij}^2$ is arbitrary and is chosen as in previous works~\cite{Schmidt2019}.
The value of $r_p$ is arbitrary and is chosen as $r_p=2.5\sigma_{cs}$ so that only satellites forming a molecule contribute to the molecule's active velocity.
However, in the strong attraction limit considered in this work, $D_s\ll\epsilon_{cs}/\gamma$, the spatial modulation shape and the cutoff value are irrelevant.

\vskip0.2cm
\noindent
{\bf Inter-molecule interactions.}
When two molecules swim towards each other, cores as well as cores and satellites of different molecules interact with each other.
Here, we assume that these interactions are conservative and due to an effective potential generated by hydrodynamic interactions.
A system with $n$ molecules evolves with dynamics \eqref{eq:molecule_dynamics} complemented with the additional inter-molecular interactions.
Core-core interactions are due to a long-range pure repulsive potential, core-satellite interactions are obtained from a long-range attractive potential, and satellite-satellite forces belonging to different molecules are negligible and simply give rise to extra repulsion.
\clearpage

Specifically, interactions between different molecules can be calculated from the total interacting potential $U_{tot}^{inter}$ which is given by:
\begin{equation}
U_{tot}^{inter}= \sum_{j<l}^n U^{inter}_{cc} (|\mathbf{x}_{c_j}-\mathbf{x}_{c_l}|) + \sum_{i=1}^{N} \sum_{j=1}^n U^{inter}_{cs} (|\mathbf{x}_{c_j}-\mathbf{x}_{i}|) \,.
\end{equation}
Here, $U^{inter}_{cc}(r)$ is the core-core repulsive potential which reads:
\begin{equation}
U_{cs}^{inter}(r)=
- 4\epsilon^{inter}_{cs} \left( \dfrac{\sigma_{cs}}{r}\right)^{0.5} \quad r<5\sigma_{cs}
\end{equation}
and zero for $r>5\sigma_{cs}$.
The constant $\epsilon^{inter}_{cs}$  sets the energy scale of this interaction. 
The potential is cut at $5\sigma_{cs}$ and is shifted to be continuous by adding the constant $C_{cs}^{inter}=-U_{cs}^{inter}(5\sigma_{cs})$. 
In addition, $U^{inter}_{cs}(r)$ is the core-satellite attracting potential chosen as:
\begin{equation}
U_{cc}^{inter}(r)=
- 4\epsilon^{inter}_{cc} \left( \dfrac{\sigma_{cc}}{r}\right)^{0.5} \quad r<5\sigma_{c}\\
\end{equation}
and zero for $r>5\sigma_{c}$.
The potential is cut at $5\sigma_{c}$ and shifted to be continuous by adding the constant $C_{cc}^{inter}=-U_{cc}^{inter}(5\sigma_{c})$. 
Finally, the constant $\epsilon^{inter}_{cc}$ determines the energy scale of this interaction. 
The total inter-molecule force acting on the satellite is given by: $\mathbf{F}^{inter}_i= \nabla_i U_{tot}^{inter}$ and the one acting on the core by: $\mathbf{F}^{inter}_{c_j}= \nabla_{c_j} U_{tot}^{inter}$.
Here, $\nabla_i$ with $i=1, ..., n$ denotes the vector derivative with respect to the satellite position, while $\nabla_{c_j}$ with $j=1, ..., N$ represents the vector derivative with respect to the core position.
Note that in this way the forces between cores and satellites of different molecules scale as $\sim 1/r^{1.5}$.
Such a choice was inspired from the shape of the effective potential generated by hydrodynamic interactions.

\vskip0.2cm
\noindent
{\textbf{Parameter choice.}
The parameters of the model are chosen according to the average experimental values, so that
the ratio between the diameters of the two particles reads $\sigma_c/\sigma_s=8$ with $\sigma_c=5.6$. 
By using the Einstein relation with a room temperature of $T=20^o$ C and
the water dynamic viscosity $\eta$, the diffusion coefficient reads 
\begin{subequations}
\begin{align}
&D_c=7.66 \times10^{-2}\mu m^2 s^{-1} \\
&D_s=6.13 \times10^{-1}\mu m^2 s^{-1} \,.
\end{align}
\end{subequations}
The active velocity of a dimer molecule with a core and a single satellite is given by:
\begin{equation}
V_1=3.4 \,\mu m s^{-1}.
\end{equation}
As a result, we can estimate the effective P\'eclet number ruling the dynamics.
The additional parameters for the numerical studies are:
\begin{subequations}
\begin{align}
&\epsilon_{cs}=5 \\
&\epsilon_d=10 \\
&\epsilon_w=1 \\
&\epsilon_{cc}^{inter}=50 \\
&\epsilon_{cs}^{inter}=1  \,.
\end{align}
\end{subequations}\\

\bibliography{Main}

\begin{thebibliography}{58}%
\makeatletter
\providecommand \@ifxundefined [1]{%
 \@ifx{#1\undefined}
}%
\providecommand \@ifnum [1]{%
 \ifnum #1\expandafter \@firstoftwo
 \else \expandafter \@secondoftwo
 \fi
}%
\providecommand \@ifx [1]{%
 \ifx #1\expandafter \@firstoftwo
 \else \expandafter \@secondoftwo
 \fi
}%
\providecommand \natexlab [1]{#1}%
\providecommand \enquote  [1]{``#1''}%
\providecommand \bibnamefont  [1]{#1}%
\providecommand \bibfnamefont [1]{#1}%
\providecommand \citenamefont [1]{#1}%
\providecommand \href@noop [0]{\@secondoftwo}%
\providecommand \href [0]{\begingroup \@sanitize@url \@href}%
\providecommand \@href[1]{\@@startlink{#1}\@@href}%
\providecommand \@@href[1]{\endgroup#1\@@endlink}%
\providecommand \@sanitize@url [0]{\catcode `\\12\catcode `\$12\catcode `\&12\catcode `\#12\catcode `\^12\catcode `\_12\catcode `\%12\relax}%
\providecommand \@@startlink[1]{}%
\providecommand \@@endlink[0]{}%
\providecommand \url  [0]{\begingroup\@sanitize@url \@url }%
\providecommand \@url [1]{\endgroup\@href {#1}{\urlprefix }}%
\providecommand \urlprefix  [0]{URL }%
\providecommand \Eprint [0]{\href }%
\providecommand \doibase [0]{https://doi.org/}%
\providecommand \selectlanguage [0]{\@gobble}%
\providecommand \bibinfo  [0]{\@secondoftwo}%
\providecommand \bibfield  [0]{\@secondoftwo}%
\providecommand \translation [1]{[#1]}%
\providecommand \BibitemOpen [0]{}%
\providecommand \bibitemStop [0]{}%
\providecommand \bibitemNoStop [0]{.\EOS\space}%
\providecommand \EOS [0]{\spacefactor3000\relax}%
\providecommand \BibitemShut  [1]{\csname bibitem#1\endcsname}%
\let\auto@bib@innerbib\@empty
\bibitem [{\citenamefont {Dreyfus}\ \emph {et~al.}(2005)\citenamefont {Dreyfus}, \citenamefont {Baudry}, \citenamefont {Roper}, \citenamefont {Fermigier}, \citenamefont {Stone},\ and\ \citenamefont {Bibette}}]{Dreyfus2005}%
  \BibitemOpen
  \bibfield  {author} {\bibinfo {author} {\bibfnamefont {R.}~\bibnamefont {Dreyfus}}, \bibinfo {author} {\bibfnamefont {J.}~\bibnamefont {Baudry}}, \bibinfo {author} {\bibfnamefont {M.~L.}\ \bibnamefont {Roper}}, \bibinfo {author} {\bibfnamefont {M.}~\bibnamefont {Fermigier}}, \bibinfo {author} {\bibfnamefont {H.~A.}\ \bibnamefont {Stone}},\ and\ \bibinfo {author} {\bibfnamefont {J.}~\bibnamefont {Bibette}},\ }\bibfield  {title} {\bibinfo {title} {Microscopic artificial swimmers},\ }\href {https://doi.org/10.1038/nature04090} {\bibfield  {journal} {\bibinfo  {journal} {Nature}\ }\textbf {\bibinfo {volume} {437}},\ \bibinfo {pages} {862} (\bibinfo {year} {2005})}\BibitemShut {NoStop}%
\bibitem [{\citenamefont {Bechinger}\ \emph {et~al.}(2016)\citenamefont {Bechinger}, \citenamefont {Leonardo}, \citenamefont {L{\"o}wen}, \citenamefont {Reichhardt}, \citenamefont {Volpe},\ and\ \citenamefont {Volpe}}]{Bechinger2016}%
  \BibitemOpen
  \bibfield  {author} {\bibinfo {author} {\bibfnamefont {C.}~\bibnamefont {Bechinger}}, \bibinfo {author} {\bibfnamefont {R.~D.}\ \bibnamefont {Leonardo}}, \bibinfo {author} {\bibfnamefont {H.}~\bibnamefont {L{\"o}wen}}, \bibinfo {author} {\bibfnamefont {C.}~\bibnamefont {Reichhardt}}, \bibinfo {author} {\bibfnamefont {G.}~\bibnamefont {Volpe}},\ and\ \bibinfo {author} {\bibfnamefont {G.}~\bibnamefont {Volpe}},\ }\bibfield  {title} {\bibinfo {title} {Active particles in complex and crowded environments},\ }\href {https://doi.org/10.1103/RevModPhys.88.045006} {\bibfield  {journal} {\bibinfo  {journal} {Rev. Mod. Phys.}\ }\textbf {\bibinfo {volume} {88}},\ \bibinfo {pages} {045006} (\bibinfo {year} {2016})}\BibitemShut {NoStop}%
\bibitem [{\citenamefont {Aubret}\ \emph {et~al.}(2018)\citenamefont {Aubret}, \citenamefont {Youssef}, \citenamefont {Sacanna},\ and\ \citenamefont {Palacci}}]{Aubret2018}%
  \BibitemOpen
  \bibfield  {author} {\bibinfo {author} {\bibfnamefont {A.}~\bibnamefont {Aubret}}, \bibinfo {author} {\bibfnamefont {M.}~\bibnamefont {Youssef}}, \bibinfo {author} {\bibfnamefont {S.}~\bibnamefont {Sacanna}},\ and\ \bibinfo {author} {\bibfnamefont {J.}~\bibnamefont {Palacci}},\ }\bibfield  {title} {\bibinfo {title} {Targeted assembly and synchronization of self-spinning microgears},\ }\href {https://doi.org/10.1038/s41567-018-0227-4} {\bibfield  {journal} {\bibinfo  {journal} {Nature Physics}\ }\textbf {\bibinfo {volume} {14}},\ \bibinfo {pages} {1114} (\bibinfo {year} {2018})}\BibitemShut {NoStop}%
\bibitem [{\citenamefont {Muiños-Landin}\ \emph {et~al.}(2021)\citenamefont {Muiños-Landin}, \citenamefont {Fischer}, \citenamefont {Holubec},\ and\ \citenamefont {Cichos}}]{Muinos2021}%
  \BibitemOpen
  \bibfield  {author} {\bibinfo {author} {\bibfnamefont {S.}~\bibnamefont {Muiños-Landin}}, \bibinfo {author} {\bibfnamefont {A.}~\bibnamefont {Fischer}}, \bibinfo {author} {\bibfnamefont {V.}~\bibnamefont {Holubec}},\ and\ \bibinfo {author} {\bibfnamefont {F.}~\bibnamefont {Cichos}},\ }\bibfield  {title} {\bibinfo {title} {Reinforcement learning with artificial microswimmers},\ }\href {https://doi.org/10.1126/scirobotics.abd9285} {\bibfield  {journal} {\bibinfo  {journal} {Science Robotics}\ }\textbf {\bibinfo {volume} {6}},\ \bibinfo {pages} {eabd9285} (\bibinfo {year} {2021})}\BibitemShut {NoStop}%
\bibitem [{\citenamefont {Zöttl}\ and\ \citenamefont {Stark}(2016)}]{Zottl2016}%
  \BibitemOpen
  \bibfield  {author} {\bibinfo {author} {\bibfnamefont {A.}~\bibnamefont {Zöttl}}\ and\ \bibinfo {author} {\bibfnamefont {H.}~\bibnamefont {Stark}},\ }\bibfield  {title} {\bibinfo {title} {Emergent behavior in active colloids},\ }\href {https://doi.org/10.1088/0953-8984/28/25/253001} {\bibfield  {journal} {\bibinfo  {journal} {J. Phys.: Condens. Matter}\ ,\ \bibinfo {pages} {253001}} (\bibinfo {year} {2016})}\BibitemShut {NoStop}%
\bibitem [{\citenamefont {Buhl}\ \emph {et~al.}(2006)\citenamefont {Buhl}, \citenamefont {Sumpter}, \citenamefont {Couzin}, \citenamefont {Hale}, \citenamefont {Despland}, \citenamefont {Miller},\ and\ \citenamefont {Simpson}}]{Buhl2006}%
  \BibitemOpen
  \bibfield  {author} {\bibinfo {author} {\bibfnamefont {J.}~\bibnamefont {Buhl}}, \bibinfo {author} {\bibfnamefont {D.~J.~T.}\ \bibnamefont {Sumpter}}, \bibinfo {author} {\bibfnamefont {I.~D.}\ \bibnamefont {Couzin}}, \bibinfo {author} {\bibfnamefont {J.~J.}\ \bibnamefont {Hale}}, \bibinfo {author} {\bibfnamefont {E.}~\bibnamefont {Despland}}, \bibinfo {author} {\bibfnamefont {E.~R.}\ \bibnamefont {Miller}},\ and\ \bibinfo {author} {\bibfnamefont {S.~J.}\ \bibnamefont {Simpson}},\ }\bibfield  {title} {\bibinfo {title} {From disorder to order in marching locusts},\ }\href {https://doi.org/10.1126/science.1125142} {\bibfield  {journal} {\bibinfo  {journal} {Science}\ }\textbf {\bibinfo {volume} {312}},\ \bibinfo {pages} {1402} (\bibinfo {year} {2006})}\BibitemShut {NoStop}%
\bibitem [{\citenamefont {Flack}\ \emph {et~al.}(2018)\citenamefont {Flack}, \citenamefont {Nagy}, \citenamefont {Fiedler}, \citenamefont {Couzin},\ and\ \citenamefont {Wikelski}}]{Flack2018}%
  \BibitemOpen
  \bibfield  {author} {\bibinfo {author} {\bibfnamefont {A.}~\bibnamefont {Flack}}, \bibinfo {author} {\bibfnamefont {M.}~\bibnamefont {Nagy}}, \bibinfo {author} {\bibfnamefont {W.}~\bibnamefont {Fiedler}}, \bibinfo {author} {\bibfnamefont {I.~D.}\ \bibnamefont {Couzin}},\ and\ \bibinfo {author} {\bibfnamefont {M.}~\bibnamefont {Wikelski}},\ }\bibfield  {title} {\bibinfo {title} {From local collective behavior to global migratory patterns in white storks},\ }\href {https://doi.org/10.1126/science.aap7781} {\bibfield  {journal} {\bibinfo  {journal} {Science}\ }\textbf {\bibinfo {volume} {360}},\ \bibinfo {pages} {911} (\bibinfo {year} {2018})}\BibitemShut {NoStop}%
\bibitem [{\citenamefont {Charlesworth}\ and\ \citenamefont {Turner}(2019)}]{Charlesworth2019}%
  \BibitemOpen
  \bibfield  {author} {\bibinfo {author} {\bibfnamefont {H.~J.}\ \bibnamefont {Charlesworth}}\ and\ \bibinfo {author} {\bibfnamefont {M.~S.}\ \bibnamefont {Turner}},\ }\bibfield  {title} {\bibinfo {title} {Intrinsically motivated collective motion},\ }\href {https://doi.org/10.1073/pnas.1822069116} {\bibfield  {journal} {\bibinfo  {journal} {PNAS}\ }\textbf {\bibinfo {volume} {116}},\ \bibinfo {pages} {15362} (\bibinfo {year} {2019})}\BibitemShut {NoStop}%
\bibitem [{\citenamefont {Kearns}(2010)}]{Kearns2010}%
  \BibitemOpen
  \bibfield  {author} {\bibinfo {author} {\bibfnamefont {D.~B.}\ \bibnamefont {Kearns}},\ }\bibfield  {title} {\bibinfo {title} {A field guide to bacterial swarming motility},\ }\href {https://doi.org/10.1038/nrmicro2405} {\bibfield  {journal} {\bibinfo  {journal} {Nat Rev Microbiol}\ }\textbf {\bibinfo {volume} {8}},\ \bibinfo {pages} {634} (\bibinfo {year} {2010})}\BibitemShut {NoStop}%
\bibitem [{\citenamefont {Schaller}\ \emph {et~al.}(2010)\citenamefont {Schaller}, \citenamefont {Weber}, \citenamefont {Semmrich}, \citenamefont {Frey},\ and\ \citenamefont {Bausch}}]{Schaller2010}%
  \BibitemOpen
  \bibfield  {author} {\bibinfo {author} {\bibfnamefont {V.}~\bibnamefont {Schaller}}, \bibinfo {author} {\bibfnamefont {C.}~\bibnamefont {Weber}}, \bibinfo {author} {\bibfnamefont {C.}~\bibnamefont {Semmrich}}, \bibinfo {author} {\bibfnamefont {E.}~\bibnamefont {Frey}},\ and\ \bibinfo {author} {\bibfnamefont {A.~R.}\ \bibnamefont {Bausch}},\ }\bibfield  {title} {\bibinfo {title} {Polar patterns of driven filaments},\ }\href {https://doi.org/10.1038/nature09312} {\bibfield  {journal} {\bibinfo  {journal} {Nature}\ }\textbf {\bibinfo {volume} {467}},\ \bibinfo {pages} {73} (\bibinfo {year} {2010})}\BibitemShut {NoStop}%
\bibitem [{\citenamefont {Sanchez}\ \emph {et~al.}(2012)\citenamefont {Sanchez}, \citenamefont {Chen}, \citenamefont {DeCamp}, \citenamefont {Heymann},\ and\ \citenamefont {Dogic}}]{Sanchez2012}%
  \BibitemOpen
  \bibfield  {author} {\bibinfo {author} {\bibfnamefont {T.}~\bibnamefont {Sanchez}}, \bibinfo {author} {\bibfnamefont {D.~T.~N.}\ \bibnamefont {Chen}}, \bibinfo {author} {\bibfnamefont {S.~J.}\ \bibnamefont {DeCamp}}, \bibinfo {author} {\bibfnamefont {M.}~\bibnamefont {Heymann}},\ and\ \bibinfo {author} {\bibfnamefont {Z.}~\bibnamefont {Dogic}},\ }\bibfield  {title} {\bibinfo {title} {Spontaneous motion in hierarchically assembled active matter},\ }\href {https://doi.org/10.1038/nature11591} {\bibfield  {journal} {\bibinfo  {journal} {Nature}\ }\textbf {\bibinfo {volume} {491}},\ \bibinfo {pages} {431} (\bibinfo {year} {2012})}\BibitemShut {NoStop}%
\bibitem [{\citenamefont {Sumino}\ \emph {et~al.}(2012)\citenamefont {Sumino}, \citenamefont {Nagai}, \citenamefont {Shitaka}, \citenamefont {Tanaka}, \citenamefont {Yoshikawa}, \citenamefont {Chaté},\ and\ \citenamefont {Oiwa}}]{Sumino2012}%
  \BibitemOpen
  \bibfield  {author} {\bibinfo {author} {\bibfnamefont {Y.}~\bibnamefont {Sumino}}, \bibinfo {author} {\bibfnamefont {K.~H.}\ \bibnamefont {Nagai}}, \bibinfo {author} {\bibfnamefont {Y.}~\bibnamefont {Shitaka}}, \bibinfo {author} {\bibfnamefont {D.}~\bibnamefont {Tanaka}}, \bibinfo {author} {\bibfnamefont {K.}~\bibnamefont {Yoshikawa}}, \bibinfo {author} {\bibfnamefont {H.}~\bibnamefont {Chaté}},\ and\ \bibinfo {author} {\bibfnamefont {K.}~\bibnamefont {Oiwa}},\ }\bibfield  {title} {\bibinfo {title} {Large-scale vortex lattice emerging from collectively moving microtubules},\ }\href {https://doi.org/10.1038/nature10874} {\bibfield  {journal} {\bibinfo  {journal} {Nature}\ }\textbf {\bibinfo {volume} {483}},\ \bibinfo {pages} {448} (\bibinfo {year} {2012})}\BibitemShut {NoStop}%
\bibitem [{\citenamefont {Henkes}\ \emph {et~al.}(2020)\citenamefont {Henkes}, \citenamefont {Kostanjevec}, \citenamefont {Collinson}, \citenamefont {Sknepnek},\ and\ \citenamefont {Bertin}}]{Henkes2020}%
  \BibitemOpen
  \bibfield  {author} {\bibinfo {author} {\bibfnamefont {S.}~\bibnamefont {Henkes}}, \bibinfo {author} {\bibfnamefont {K.}~\bibnamefont {Kostanjevec}}, \bibinfo {author} {\bibfnamefont {J.~M.}\ \bibnamefont {Collinson}}, \bibinfo {author} {\bibfnamefont {R.}~\bibnamefont {Sknepnek}},\ and\ \bibinfo {author} {\bibfnamefont {E.}~\bibnamefont {Bertin}},\ }\bibfield  {title} {\bibinfo {title} {Dense active matter model of motion patterns in confluent cell monolayers},\ }\href@noop {} {\bibfield  {journal} {\bibinfo  {journal} {Nat Commun}\ }\textbf {\bibinfo {volume} {11}} (\bibinfo {year} {2020})}\BibitemShut {NoStop}%
\bibitem [{\citenamefont {Stramer}\ and\ \citenamefont {Mayor}(2017)}]{Stramer2017}%
  \BibitemOpen
  \bibfield  {author} {\bibinfo {author} {\bibfnamefont {B.}~\bibnamefont {Stramer}}\ and\ \bibinfo {author} {\bibfnamefont {R.}~\bibnamefont {Mayor}},\ }\bibfield  {title} {\bibinfo {title} {Mechanisms and in vivo functions of contact inhibition of locomotion},\ }\href {https://doi.org/10.1038/nrm.2016.118} {\bibfield  {journal} {\bibinfo  {journal} {Nat Rev Mol Cell Biol}\ }\textbf {\bibinfo {volume} {18}},\ \bibinfo {pages} {43} (\bibinfo {year} {2017})}\BibitemShut {NoStop}%
\bibitem [{\citenamefont {Singh}\ \emph {et~al.}(2021)\citenamefont {Singh}, \citenamefont {Pagulayan}, \citenamefont {Camley},\ and\ \citenamefont {Nain}}]{Singh2021}%
  \BibitemOpen
  \bibfield  {author} {\bibinfo {author} {\bibfnamefont {J.}~\bibnamefont {Singh}}, \bibinfo {author} {\bibfnamefont {A.}~\bibnamefont {Pagulayan}}, \bibinfo {author} {\bibfnamefont {B.~A.}\ \bibnamefont {Camley}},\ and\ \bibinfo {author} {\bibfnamefont {A.~S.}\ \bibnamefont {Nain}},\ }\bibfield  {title} {\bibinfo {title} {Rules of contact inhibition of locomotion for cells on suspended nanofibers},\ }\href@noop {} {\bibfield  {journal} {\bibinfo  {journal} {PNAS}\ }\textbf {\bibinfo {volume} {118}} (\bibinfo {year} {2021})}\BibitemShut {NoStop}%
\bibitem [{\citenamefont {Lämmermann}\ \emph {et~al.}(2008)\citenamefont {Lämmermann}, \citenamefont {Bader}, \citenamefont {Monkley}, \citenamefont {Worbs}, \citenamefont {Wedlich-Söldner}, \citenamefont {Hirsch}, \citenamefont {Keller}, \citenamefont {Förster}, \citenamefont {Critchley}, \citenamefont {Fässler},\ and\ \citenamefont {Sixt}}]{Lammermann2008}%
  \BibitemOpen
  \bibfield  {author} {\bibinfo {author} {\bibfnamefont {T.}~\bibnamefont {Lämmermann}}, \bibinfo {author} {\bibfnamefont {B.~L.}\ \bibnamefont {Bader}}, \bibinfo {author} {\bibfnamefont {S.~J.}\ \bibnamefont {Monkley}}, \bibinfo {author} {\bibfnamefont {T.}~\bibnamefont {Worbs}}, \bibinfo {author} {\bibfnamefont {R.}~\bibnamefont {Wedlich-Söldner}}, \bibinfo {author} {\bibfnamefont {K.}~\bibnamefont {Hirsch}}, \bibinfo {author} {\bibfnamefont {M.}~\bibnamefont {Keller}}, \bibinfo {author} {\bibfnamefont {R.}~\bibnamefont {Förster}}, \bibinfo {author} {\bibfnamefont {D.~R.}\ \bibnamefont {Critchley}}, \bibinfo {author} {\bibfnamefont {R.}~\bibnamefont {Fässler}},\ and\ \bibinfo {author} {\bibfnamefont {M.}~\bibnamefont {Sixt}},\ }\bibfield  {title} {\bibinfo {title} {Rapid leukocyte migration by integrin-independent flowing and squeezing},\ }\href {https://doi.org/10.1038/nature06887} {\bibfield  {journal} {\bibinfo  {journal} {Nature}\ }\textbf {\bibinfo {volume} {453}},\ \bibinfo {pages} {51} (\bibinfo
  {year} {2008})}\BibitemShut {NoStop}%
\bibitem [{\citenamefont {Carmona-Fontaine}\ \emph {et~al.}(2008)\citenamefont {Carmona-Fontaine}, \citenamefont {Matthews}, \citenamefont {Kuriyama}, \citenamefont {Moreno}, \citenamefont {Dunn}, \citenamefont {Parsons}, \citenamefont {Stern},\ and\ \citenamefont {Mayor}}]{Carmona2008}%
  \BibitemOpen
  \bibfield  {author} {\bibinfo {author} {\bibfnamefont {C.}~\bibnamefont {Carmona-Fontaine}}, \bibinfo {author} {\bibfnamefont {H.~K.}\ \bibnamefont {Matthews}}, \bibinfo {author} {\bibfnamefont {S.}~\bibnamefont {Kuriyama}}, \bibinfo {author} {\bibfnamefont {M.}~\bibnamefont {Moreno}}, \bibinfo {author} {\bibfnamefont {G.~A.}\ \bibnamefont {Dunn}}, \bibinfo {author} {\bibfnamefont {M.}~\bibnamefont {Parsons}}, \bibinfo {author} {\bibfnamefont {C.~D.}\ \bibnamefont {Stern}},\ and\ \bibinfo {author} {\bibfnamefont {R.}~\bibnamefont {Mayor}},\ }\bibfield  {title} {\bibinfo {title} {Contact inhibition of locomotion in vivo controls neural crest directional migration},\ }\href {https://doi.org/10.1038/nature07441} {\bibfield  {journal} {\bibinfo  {journal} {Nature}\ }\textbf {\bibinfo {volume} {456}},\ \bibinfo {pages} {957} (\bibinfo {year} {2008})}\BibitemShut {NoStop}%
\bibitem [{\citenamefont {Palacci}\ \emph {et~al.}(2013)\citenamefont {Palacci}, \citenamefont {Sacanna}, \citenamefont {Steinberg}, \citenamefont {Pine},\ and\ \citenamefont {Chaikin}}]{Palacci2013}%
  \BibitemOpen
  \bibfield  {author} {\bibinfo {author} {\bibfnamefont {J.}~\bibnamefont {Palacci}}, \bibinfo {author} {\bibfnamefont {S.}~\bibnamefont {Sacanna}}, \bibinfo {author} {\bibfnamefont {A.~P.}\ \bibnamefont {Steinberg}}, \bibinfo {author} {\bibfnamefont {D.~J.}\ \bibnamefont {Pine}},\ and\ \bibinfo {author} {\bibfnamefont {P.~M.}\ \bibnamefont {Chaikin}},\ }\bibfield  {title} {\bibinfo {title} {Living crystals of light-activated colloidal surfers},\ }\href {https://doi.org/10.1126/science.1230020} {\bibfield  {journal} {\bibinfo  {journal} {Science}\ }\textbf {\bibinfo {volume} {339}},\ \bibinfo {pages} {936} (\bibinfo {year} {2013})}\BibitemShut {NoStop}%
\bibitem [{\citenamefont {Theurkauff}\ \emph {et~al.}(2012)\citenamefont {Theurkauff}, \citenamefont {Cottin-Bizonne}, \citenamefont {Palacci}, \citenamefont {Ybert},\ and\ \citenamefont {Bocquet}}]{Theurkauff2012}%
  \BibitemOpen
  \bibfield  {author} {\bibinfo {author} {\bibfnamefont {I.}~\bibnamefont {Theurkauff}}, \bibinfo {author} {\bibfnamefont {C.}~\bibnamefont {Cottin-Bizonne}}, \bibinfo {author} {\bibfnamefont {J.}~\bibnamefont {Palacci}}, \bibinfo {author} {\bibfnamefont {C.}~\bibnamefont {Ybert}},\ and\ \bibinfo {author} {\bibfnamefont {L.}~\bibnamefont {Bocquet}},\ }\bibfield  {title} {\bibinfo {title} {Dynamic clustering in active colloidal suspensions with chemical signaling},\ }\href {https://doi.org/10.1103/PhysRevLett.108.268303} {\bibfield  {journal} {\bibinfo  {journal} {Phys. Rev. Lett.}\ }\textbf {\bibinfo {volume} {108}},\ \bibinfo {pages} {268303} (\bibinfo {year} {2012})}\BibitemShut {NoStop}%
\bibitem [{\citenamefont {Buttinoni}\ \emph {et~al.}(2013)\citenamefont {Buttinoni}, \citenamefont {Bialk\'e}, \citenamefont {K\"ummel}, \citenamefont {L\"owen}, \citenamefont {Bechinger},\ and\ \citenamefont {Speck}}]{Buttinoni2013}%
  \BibitemOpen
  \bibfield  {author} {\bibinfo {author} {\bibfnamefont {I.}~\bibnamefont {Buttinoni}}, \bibinfo {author} {\bibfnamefont {J.}~\bibnamefont {Bialk\'e}}, \bibinfo {author} {\bibfnamefont {F.}~\bibnamefont {K\"ummel}}, \bibinfo {author} {\bibfnamefont {H.}~\bibnamefont {L\"owen}}, \bibinfo {author} {\bibfnamefont {C.}~\bibnamefont {Bechinger}},\ and\ \bibinfo {author} {\bibfnamefont {T.}~\bibnamefont {Speck}},\ }\bibfield  {title} {\bibinfo {title} {Dynamical clustering and phase separation in suspensions of self-propelled colloidal particles},\ }\href {https://doi.org/10.1103/PhysRevLett.110.238301} {\bibfield  {journal} {\bibinfo  {journal} {Phys. Rev. Lett.}\ }\textbf {\bibinfo {volume} {110}},\ \bibinfo {pages} {238301} (\bibinfo {year} {2013})}\BibitemShut {NoStop}%
\bibitem [{\citenamefont {Ginot}\ \emph {et~al.}(2018)\citenamefont {Ginot}, \citenamefont {Theurkauff}, \citenamefont {Detcheverry}, \citenamefont {Ybert},\ and\ \citenamefont {Cottin-Bizonne}}]{Ginot2018}%
  \BibitemOpen
  \bibfield  {author} {\bibinfo {author} {\bibfnamefont {F.}~\bibnamefont {Ginot}}, \bibinfo {author} {\bibfnamefont {I.}~\bibnamefont {Theurkauff}}, \bibinfo {author} {\bibfnamefont {F.}~\bibnamefont {Detcheverry}}, \bibinfo {author} {\bibfnamefont {C.}~\bibnamefont {Ybert}},\ and\ \bibinfo {author} {\bibfnamefont {C.}~\bibnamefont {Cottin-Bizonne}},\ }\bibfield  {title} {\bibinfo {title} {Aggregation-fragmentation and individual dynamics of active clusters},\ }\href@noop {} {\bibfield  {journal} {\bibinfo  {journal} {Nat Commun}\ }\textbf {\bibinfo {volume} {9}},\ \bibinfo {pages} {696} (\bibinfo {year} {2018})}\BibitemShut {NoStop}%
\bibitem [{\citenamefont {Bricard}\ \emph {et~al.}(2013)\citenamefont {Bricard}, \citenamefont {Caussin}, \citenamefont {Desreumaux}, \citenamefont {Dauchot},\ and\ \citenamefont {Bartolo}}]{Bricard2013}%
  \BibitemOpen
  \bibfield  {author} {\bibinfo {author} {\bibfnamefont {A.}~\bibnamefont {Bricard}}, \bibinfo {author} {\bibfnamefont {J.-B.}\ \bibnamefont {Caussin}}, \bibinfo {author} {\bibfnamefont {N.}~\bibnamefont {Desreumaux}}, \bibinfo {author} {\bibfnamefont {O.}~\bibnamefont {Dauchot}},\ and\ \bibinfo {author} {\bibfnamefont {D.}~\bibnamefont {Bartolo}},\ }\bibfield  {title} {\bibinfo {title} {Emergence of macroscopic directed motion in populations of motile colloids},\ }\href {https://doi.org/10.1038/nature12673} {\bibfield  {journal} {\bibinfo  {journal} {Nature}\ }\textbf {\bibinfo {volume} {503}},\ \bibinfo {pages} {95} (\bibinfo {year} {2013})}\BibitemShut {NoStop}%
\bibitem [{\citenamefont {Digregorio}\ \emph {et~al.}(2018)\citenamefont {Digregorio}, \citenamefont {Levis}, \citenamefont {Suma}, \citenamefont {Cugliandolo}, \citenamefont {Gonnella},\ and\ \citenamefont {Pagonabarraga}}]{Digregorio2018}%
  \BibitemOpen
  \bibfield  {author} {\bibinfo {author} {\bibfnamefont {P.}~\bibnamefont {Digregorio}}, \bibinfo {author} {\bibfnamefont {D.}~\bibnamefont {Levis}}, \bibinfo {author} {\bibfnamefont {A.}~\bibnamefont {Suma}}, \bibinfo {author} {\bibfnamefont {L.~F.}\ \bibnamefont {Cugliandolo}}, \bibinfo {author} {\bibfnamefont {G.}~\bibnamefont {Gonnella}},\ and\ \bibinfo {author} {\bibfnamefont {I.}~\bibnamefont {Pagonabarraga}},\ }\bibfield  {title} {\bibinfo {title} {Full phase diagram of active brownian disks: From melting to motility-induced phase separation},\ }\href {https://doi.org/10.1103/PhysRevLett.121.098003} {\bibfield  {journal} {\bibinfo  {journal} {Phys Rev Lett}\ }\textbf {\bibinfo {volume} {121}},\ \bibinfo {pages} {098003} (\bibinfo {year} {2018})}\BibitemShut {NoStop}%
\bibitem [{\citenamefont {Bricard}\ \emph {et~al.}(2015)\citenamefont {Bricard}, \citenamefont {Caussin}, \citenamefont {Das}, \citenamefont {Savoie}, \citenamefont {Chikkadi}, \citenamefont {Shitara}, \citenamefont {Chepizhko}, \citenamefont {Peruani}, \citenamefont {Saintillan},\ and\ \citenamefont {Bartolo}}]{Bricard2015}%
  \BibitemOpen
  \bibfield  {author} {\bibinfo {author} {\bibfnamefont {A.}~\bibnamefont {Bricard}}, \bibinfo {author} {\bibfnamefont {J.-B.}\ \bibnamefont {Caussin}}, \bibinfo {author} {\bibfnamefont {D.}~\bibnamefont {Das}}, \bibinfo {author} {\bibfnamefont {C.}~\bibnamefont {Savoie}}, \bibinfo {author} {\bibfnamefont {V.}~\bibnamefont {Chikkadi}}, \bibinfo {author} {\bibfnamefont {K.}~\bibnamefont {Shitara}}, \bibinfo {author} {\bibfnamefont {O.}~\bibnamefont {Chepizhko}}, \bibinfo {author} {\bibfnamefont {F.}~\bibnamefont {Peruani}}, \bibinfo {author} {\bibfnamefont {D.}~\bibnamefont {Saintillan}},\ and\ \bibinfo {author} {\bibfnamefont {D.}~\bibnamefont {Bartolo}},\ }\bibfield  {title} {\bibinfo {title} {Emergent vortices in populations of colloidal rollers},\ }\href@noop {} {\bibfield  {journal} {\bibinfo  {journal} {Nat Commun}\ }\textbf {\bibinfo {volume} {6}},\ \bibinfo {pages} {7470} (\bibinfo {year} {2015})}\BibitemShut {NoStop}%
\bibitem [{\citenamefont {Yan}\ \emph {et~al.}(2016)\citenamefont {Yan}, \citenamefont {Han}, \citenamefont {Zhang}, \citenamefont {Xu}, \citenamefont {Luijten},\ and\ \citenamefont {Granick}}]{Yan2016}%
  \BibitemOpen
  \bibfield  {author} {\bibinfo {author} {\bibfnamefont {J.}~\bibnamefont {Yan}}, \bibinfo {author} {\bibfnamefont {M.}~\bibnamefont {Han}}, \bibinfo {author} {\bibfnamefont {J.}~\bibnamefont {Zhang}}, \bibinfo {author} {\bibfnamefont {C.}~\bibnamefont {Xu}}, \bibinfo {author} {\bibfnamefont {E.}~\bibnamefont {Luijten}},\ and\ \bibinfo {author} {\bibfnamefont {S.}~\bibnamefont {Granick}},\ }\bibfield  {title} {\bibinfo {title} {Reconfiguring active particles by electrostatic imbalance},\ }\href {https://doi.org/10.1038/nmat4696} {\bibfield  {journal} {\bibinfo  {journal} {Nature Materials}\ }\textbf {\bibinfo {volume} {15}},\ \bibinfo {pages} {1095} (\bibinfo {year} {2016})}\BibitemShut {NoStop}%
\bibitem [{\citenamefont {Kaiser}\ \emph {et~al.}(2017)\citenamefont {Kaiser}, \citenamefont {Snezhko},\ and\ \citenamefont {Aranson}}]{Kaiser2017}%
  \BibitemOpen
  \bibfield  {author} {\bibinfo {author} {\bibfnamefont {A.}~\bibnamefont {Kaiser}}, \bibinfo {author} {\bibfnamefont {A.}~\bibnamefont {Snezhko}},\ and\ \bibinfo {author} {\bibfnamefont {I.~S.}\ \bibnamefont {Aranson}},\ }\bibfield  {title} {\bibinfo {title} {Flocking ferromagnetic colloids},\ }\href {https://doi.org/10.1126/sciadv.1601469} {\bibfield  {journal} {\bibinfo  {journal} {Sci. Adv.}\ }\textbf {\bibinfo {volume} {3}},\ \bibinfo {pages} {e1601469} (\bibinfo {year} {2017})}\BibitemShut {NoStop}%
\bibitem [{\citenamefont {Karani}\ \emph {et~al.}(2019)\citenamefont {Karani}, \citenamefont {Pradillo},\ and\ \citenamefont {Vlahovska}}]{Karani2019}%
  \BibitemOpen
  \bibfield  {author} {\bibinfo {author} {\bibfnamefont {H.}~\bibnamefont {Karani}}, \bibinfo {author} {\bibfnamefont {G.~E.}\ \bibnamefont {Pradillo}},\ and\ \bibinfo {author} {\bibfnamefont {P.~M.}\ \bibnamefont {Vlahovska}},\ }\bibfield  {title} {\bibinfo {title} {Tuning the random walk of active colloids: From individual run-and-tumble to dynamic clustering},\ }\href {https://doi.org/10.1103/PhysRevLett.123.208002} {\bibfield  {journal} {\bibinfo  {journal} {Physical Review Letters}\ }\textbf {\bibinfo {volume} {123}},\ \bibinfo {pages} {208002} (\bibinfo {year} {2019})}\BibitemShut {NoStop}%
\bibitem [{\citenamefont {Soni}\ \emph {et~al.}(2019)\citenamefont {Soni}, \citenamefont {Bililign}, \citenamefont {Magkiriadou}, \citenamefont {Sacanna}, \citenamefont {Bartolo}, \citenamefont {Shelley},\ and\ \citenamefont {Irvine}}]{Soni2019}%
  \BibitemOpen
  \bibfield  {author} {\bibinfo {author} {\bibfnamefont {V.}~\bibnamefont {Soni}}, \bibinfo {author} {\bibfnamefont {E.~S.}\ \bibnamefont {Bililign}}, \bibinfo {author} {\bibfnamefont {S.}~\bibnamefont {Magkiriadou}}, \bibinfo {author} {\bibfnamefont {S.}~\bibnamefont {Sacanna}}, \bibinfo {author} {\bibfnamefont {D.}~\bibnamefont {Bartolo}}, \bibinfo {author} {\bibfnamefont {M.~J.}\ \bibnamefont {Shelley}},\ and\ \bibinfo {author} {\bibfnamefont {W.~T.~M.}\ \bibnamefont {Irvine}},\ }\bibfield  {title} {\bibinfo {title} {The odd free surface flows of a colloidal chiral fluid},\ }\href {https://doi.org/10.1038/s41567-019-0603-8} {\bibfield  {journal} {\bibinfo  {journal} {Nat Phys}\ ,\ \bibinfo {pages} {1188}} (\bibinfo {year} {2019})}\BibitemShut {NoStop}%
\bibitem [{\citenamefont {van~der Linden}\ \emph {et~al.}(2019)\citenamefont {van~der Linden}, \citenamefont {Alexander}, \citenamefont {Aarts},\ and\ \citenamefont {Dauchot}}]{Linden2019}%
  \BibitemOpen
  \bibfield  {author} {\bibinfo {author} {\bibfnamefont {M.~N.}\ \bibnamefont {van~der Linden}}, \bibinfo {author} {\bibfnamefont {L.~C.}\ \bibnamefont {Alexander}}, \bibinfo {author} {\bibfnamefont {D.~G. A.~L.}\ \bibnamefont {Aarts}},\ and\ \bibinfo {author} {\bibfnamefont {O.}~\bibnamefont {Dauchot}},\ }\bibfield  {title} {\bibinfo {title} {Interrupted motility induced phase separation in aligning active colloids},\ }\href {https://doi.org/10.1103/PhysRevLett.123.098001} {\bibfield  {journal} {\bibinfo  {journal} {Phys. Rev. Lett.}\ }\textbf {\bibinfo {volume} {123}},\ \bibinfo {pages} {098001} (\bibinfo {year} {2019})}\BibitemShut {NoStop}%
\bibitem [{\citenamefont {Geyer}\ \emph {et~al.}(2019)\citenamefont {Geyer}, \citenamefont {Martin}, \citenamefont {Tailleur},\ and\ \citenamefont {Bartolo}}]{Geyer2019}%
  \BibitemOpen
  \bibfield  {author} {\bibinfo {author} {\bibfnamefont {D.}~\bibnamefont {Geyer}}, \bibinfo {author} {\bibfnamefont {D.}~\bibnamefont {Martin}}, \bibinfo {author} {\bibfnamefont {J.}~\bibnamefont {Tailleur}},\ and\ \bibinfo {author} {\bibfnamefont {D.}~\bibnamefont {Bartolo}},\ }\bibfield  {title} {\bibinfo {title} {Freezing a flock: Motility-induced phase separation in polar active liquids},\ }\href {https://doi.org/10.1103/PhysRevX.9.031043} {\bibfield  {journal} {\bibinfo  {journal} {Phys. Rev. X}\ }\textbf {\bibinfo {volume} {9}},\ \bibinfo {pages} {031043} (\bibinfo {year} {2019})}\BibitemShut {NoStop}%
\bibitem [{\citenamefont {Ketzetzi}\ \emph {et~al.}(2022)\citenamefont {Ketzetzi}, \citenamefont {Rinaldin}, \citenamefont {Dröge}, \citenamefont {de~Graaf},\ and\ \citenamefont {Kraft}}]{Ketzetzi2022}%
  \BibitemOpen
  \bibfield  {author} {\bibinfo {author} {\bibfnamefont {S.}~\bibnamefont {Ketzetzi}}, \bibinfo {author} {\bibfnamefont {M.}~\bibnamefont {Rinaldin}}, \bibinfo {author} {\bibfnamefont {P.}~\bibnamefont {Dröge}}, \bibinfo {author} {\bibfnamefont {J.}~\bibnamefont {de~Graaf}},\ and\ \bibinfo {author} {\bibfnamefont {D.~J.}\ \bibnamefont {Kraft}},\ }\bibfield  {title} {\bibinfo {title} {Activity-induced interactions and cooperation of artificial microswimmers in one-dimensional environments},\ }\href {https://doi.org/10.1038/s41467-022-29430-1} {\bibfield  {journal} {\bibinfo  {journal} {Nat Commun}\ }\textbf {\bibinfo {volume} {13}},\ \bibinfo {pages} {1772} (\bibinfo {year} {2022})}\BibitemShut {NoStop}%
\bibitem [{\citenamefont {Zion}\ \emph {et~al.}(2022)\citenamefont {Zion}, \citenamefont {Caba}, \citenamefont {Modin},\ and\ \citenamefont {Chaikin}}]{Zion2022}%
  \BibitemOpen
  \bibfield  {author} {\bibinfo {author} {\bibfnamefont {M.~Y.~B.}\ \bibnamefont {Zion}}, \bibinfo {author} {\bibfnamefont {Y.}~\bibnamefont {Caba}}, \bibinfo {author} {\bibfnamefont {A.}~\bibnamefont {Modin}},\ and\ \bibinfo {author} {\bibfnamefont {P.~M.}\ \bibnamefont {Chaikin}},\ }\bibfield  {title} {\bibinfo {title} {Cooperation in a fluid swarm of fuel-free micro-swimmers},\ }\href@noop {} {\bibfield  {journal} {\bibinfo  {journal} {Nat Commun}\ }\textbf {\bibinfo {volume} {13}} (\bibinfo {year} {2022})}\BibitemShut {NoStop}%
\bibitem [{\citenamefont {Das}\ \emph {et~al.}(2024)\citenamefont {Das}, \citenamefont {Ciarchi}, \citenamefont {Zhou}, \citenamefont {Yan}, \citenamefont {Zhang},\ and\ \citenamefont {Alert}}]{Das2024}%
  \BibitemOpen
  \bibfield  {author} {\bibinfo {author} {\bibfnamefont {S.}~\bibnamefont {Das}}, \bibinfo {author} {\bibfnamefont {M.}~\bibnamefont {Ciarchi}}, \bibinfo {author} {\bibfnamefont {Z.}~\bibnamefont {Zhou}}, \bibinfo {author} {\bibfnamefont {J.}~\bibnamefont {Yan}}, \bibinfo {author} {\bibfnamefont {J.}~\bibnamefont {Zhang}},\ and\ \bibinfo {author} {\bibfnamefont {R.}~\bibnamefont {Alert}},\ }\bibfield  {title} {\bibinfo {title} {Flocking by turning away},\ }\href {https://doi.org/https://doi.org/10.1103/PhysRevX.14.031008} {\bibfield  {journal} {\bibinfo  {journal} {Phys. Rev. X}\ }\textbf {\bibinfo {volume} {14}},\ \bibinfo {pages} {031008} (\bibinfo {year} {2024})}\BibitemShut {NoStop}%
\bibitem [{\citenamefont {McCleland}\ \emph {et~al.}(2009)\citenamefont {McCleland}, \citenamefont {Farrell},\ and\ \citenamefont {O'Farrell}}]{Cleland2009}%
  \BibitemOpen
  \bibfield  {author} {\bibinfo {author} {\bibfnamefont {M.~L.}\ \bibnamefont {McCleland}}, \bibinfo {author} {\bibfnamefont {J.~A.}\ \bibnamefont {Farrell}},\ and\ \bibinfo {author} {\bibfnamefont {P.~H.}\ \bibnamefont {O'Farrell}},\ }\bibfield  {title} {\bibinfo {title} {Influence of cyclin type and dose on mitotic entry and progression in the early drosophila embryo},\ }\href {https://doi.org/10.1083/jcb.200810012} {\bibfield  {journal} {\bibinfo  {journal} {J Cell Biol}\ }\textbf {\bibinfo {volume} {184}},\ \bibinfo {pages} {639} (\bibinfo {year} {2009})}\BibitemShut {NoStop}%
\bibitem [{\citenamefont {Huang}\ \emph {et~al.}(2021)\citenamefont {Huang}, \citenamefont {Hu}, \citenamefont {Yang}, \citenamefont {Liu},\ and\ \citenamefont {Zhang}}]{Huang2021}%
  \BibitemOpen
  \bibfield  {author} {\bibinfo {author} {\bibfnamefont {M.}~\bibnamefont {Huang}}, \bibinfo {author} {\bibfnamefont {W.}~\bibnamefont {Hu}}, \bibinfo {author} {\bibfnamefont {S.}~\bibnamefont {Yang}}, \bibinfo {author} {\bibfnamefont {Q.-X.}\ \bibnamefont {Liu}},\ and\ \bibinfo {author} {\bibfnamefont {H.~P.}\ \bibnamefont {Zhang}},\ }\bibfield  {title} {\bibinfo {title} {Circular swimming motility and disordered hyperuniform state in an algae system},\ }\href {https://doi.org/10.1073/pnas.2100493118} {\bibfield  {journal} {\bibinfo  {journal} {PNAS}\ }\textbf {\bibinfo {volume} {118}},\ \bibinfo {pages} {e2100493118} (\bibinfo {year} {2021})}\BibitemShut {NoStop}%
\bibitem [{\citenamefont {Angelini}\ \emph {et~al.}(2011)\citenamefont {Angelini}, \citenamefont {Hannezo}, \citenamefont {Trepat}, \citenamefont {Marquez}, \citenamefont {Fredberg},\ and\ \citenamefont {Weitz}}]{Angelini2011}%
  \BibitemOpen
  \bibfield  {author} {\bibinfo {author} {\bibfnamefont {T.~E.}\ \bibnamefont {Angelini}}, \bibinfo {author} {\bibfnamefont {E.}~\bibnamefont {Hannezo}}, \bibinfo {author} {\bibfnamefont {X.}~\bibnamefont {Trepat}}, \bibinfo {author} {\bibfnamefont {M.}~\bibnamefont {Marquez}}, \bibinfo {author} {\bibfnamefont {J.~J.}\ \bibnamefont {Fredberg}},\ and\ \bibinfo {author} {\bibfnamefont {D.~A.}\ \bibnamefont {Weitz}},\ }\bibfield  {title} {\bibinfo {title} {Glass-like dynamics of collective cell migration},\ }\href {https://doi.org/10.1073/pnas.1010059108} {\bibfield  {journal} {\bibinfo  {journal} {PNAS}\ }\textbf {\bibinfo {volume} {108}},\ \bibinfo {pages} {4714} (\bibinfo {year} {2011})}\BibitemShut {NoStop}%
\bibitem [{\citenamefont {Kaiser}\ \emph {et~al.}(2018)\citenamefont {Kaiser}, \citenamefont {Lv}, \citenamefont {Rodrigues}, \citenamefont {Rosenbaum}, \citenamefont {Aspelmeier}, \citenamefont {Großhans},\ and\ \citenamefont {Alim}}]{Kaiser2018}%
  \BibitemOpen
  \bibfield  {author} {\bibinfo {author} {\bibfnamefont {F.}~\bibnamefont {Kaiser}}, \bibinfo {author} {\bibfnamefont {Z.}~\bibnamefont {Lv}}, \bibinfo {author} {\bibfnamefont {D.~M.}\ \bibnamefont {Rodrigues}}, \bibinfo {author} {\bibfnamefont {J.}~\bibnamefont {Rosenbaum}}, \bibinfo {author} {\bibfnamefont {T.}~\bibnamefont {Aspelmeier}}, \bibinfo {author} {\bibfnamefont {J.}~\bibnamefont {Großhans}},\ and\ \bibinfo {author} {\bibfnamefont {K.}~\bibnamefont {Alim}},\ }\bibfield  {title} {\bibinfo {title} {Mechanical model of nuclei ordering in drosophila embryos reveals dilution of stochastic forces},\ }\href {https://doi.org/https://doi.org/10.1016/j.bpj.2018.02.018} {\bibfield  {journal} {\bibinfo  {journal} {Biophysical Journal}\ }\textbf {\bibinfo {volume} {114}},\ \bibinfo {pages} {1730} (\bibinfo {year} {2018})}\BibitemShut {NoStop}%
\bibitem [{\citenamefont {Shields}\ and\ \citenamefont {Velev}(2017)}]{Shields2017}%
  \BibitemOpen
  \bibfield  {author} {\bibinfo {author} {\bibfnamefont {C.~W.}\ \bibnamefont {Shields}}\ and\ \bibinfo {author} {\bibfnamefont {O.~D.}\ \bibnamefont {Velev}},\ }\bibfield  {title} {\bibinfo {title} {The evolution of active particles: Toward externally powered self-propelling and self-reconfiguring particle systems},\ }\href {https://doi.org/10.1016/j.chempr.2017.09.006} {\bibfield  {journal} {\bibinfo  {journal} {Chem}\ }\textbf {\bibinfo {volume} {3}},\ \bibinfo {pages} {539} (\bibinfo {year} {2017})}\BibitemShut {NoStop}%
\bibitem [{\citenamefont {Zhang}\ \emph {et~al.}(2016)\citenamefont {Zhang}, \citenamefont {Yan},\ and\ \citenamefont {Granick}}]{Zhang2016}%
  \BibitemOpen
  \bibfield  {author} {\bibinfo {author} {\bibfnamefont {J.}~\bibnamefont {Zhang}}, \bibinfo {author} {\bibfnamefont {J.}~\bibnamefont {Yan}},\ and\ \bibinfo {author} {\bibfnamefont {S.}~\bibnamefont {Granick}},\ }\bibfield  {title} {\bibinfo {title} {Directed self-assembly pathways of active colloidal clusters},\ }\href {https://doi.org/10.1002/anie.201509978} {\bibfield  {journal} {\bibinfo  {journal} {Angew. Chem. Int. Ed.}\ }\textbf {\bibinfo {volume} {55}},\ \bibinfo {pages} {5166} (\bibinfo {year} {2016})}\BibitemShut {NoStop}%
\bibitem [{\citenamefont {Han}\ \emph {et~al.}(2018)\citenamefont {Han}, \citenamefont {Shields},\ and\ \citenamefont {Velev}}]{Han2018}%
  \BibitemOpen
  \bibfield  {author} {\bibinfo {author} {\bibfnamefont {K.}~\bibnamefont {Han}}, \bibinfo {author} {\bibfnamefont {C.~W.}\ \bibnamefont {Shields}},\ and\ \bibinfo {author} {\bibfnamefont {O.~D.}\ \bibnamefont {Velev}},\ }\bibfield  {title} {\bibinfo {title} {Engineering of self-propelling microbots and microdevices powered by magnetic and electric fields},\ }\href {https://doi.org/10.1002/adfm.201705953} {\bibfield  {journal} {\bibinfo  {journal} {Adv. Funct. Mater.}\ }\textbf {\bibinfo {volume} {28}},\ \bibinfo {pages} {1705953} (\bibinfo {year} {2018})}\BibitemShut {NoStop}%
\bibitem [{\citenamefont {Alvarez}\ \emph {et~al.}(2021)\citenamefont {Alvarez}, \citenamefont {Fernandez-Rodriguez}, \citenamefont {Alegria}, \citenamefont {Arrese-Igor}, \citenamefont {Zhao}, \citenamefont {Kröger},\ and\ \citenamefont {Isa}}]{Alvarez2021}%
  \BibitemOpen
  \bibfield  {author} {\bibinfo {author} {\bibfnamefont {L.}~\bibnamefont {Alvarez}}, \bibinfo {author} {\bibfnamefont {M.~A.}\ \bibnamefont {Fernandez-Rodriguez}}, \bibinfo {author} {\bibfnamefont {A.}~\bibnamefont {Alegria}}, \bibinfo {author} {\bibfnamefont {S.}~\bibnamefont {Arrese-Igor}}, \bibinfo {author} {\bibfnamefont {K.}~\bibnamefont {Zhao}}, \bibinfo {author} {\bibfnamefont {M.}~\bibnamefont {Kröger}},\ and\ \bibinfo {author} {\bibfnamefont {L.}~\bibnamefont {Isa}},\ }\bibfield  {title} {\bibinfo {title} {Reconfigurable artificial microswimmers with internal feedback},\ }\href {https://doi.org/10.1038/s41467-021-25108-2} {\bibfield  {journal} {\bibinfo  {journal} {Nat Commun}\ }\textbf {\bibinfo {volume} {12}},\ \bibinfo {pages} {4762} (\bibinfo {year} {2021})}\BibitemShut {NoStop}%
\bibitem [{\citenamefont {Wang}\ \emph {et~al.}(2020)\citenamefont {Wang}, \citenamefont {Wang}, \citenamefont {Li}, \citenamefont {Tian},\ and\ \citenamefont {Wang}}]{Wang2020}%
  \BibitemOpen
  \bibfield  {author} {\bibinfo {author} {\bibfnamefont {Z.}~\bibnamefont {Wang}}, \bibinfo {author} {\bibfnamefont {Z.}~\bibnamefont {Wang}}, \bibinfo {author} {\bibfnamefont {J.}~\bibnamefont {Li}}, \bibinfo {author} {\bibfnamefont {C.}~\bibnamefont {Tian}},\ and\ \bibinfo {author} {\bibfnamefont {Y.}~\bibnamefont {Wang}},\ }\bibfield  {title} {\bibinfo {title} {Active colloidal molecules assembled via selective and directional bonds},\ }\href {https://doi.org/10.1038/s41467-020-16506-z} {\bibfield  {journal} {\bibinfo  {journal} {Nat Comm}\ }\textbf {\bibinfo {volume} {11}},\ \bibinfo {pages} {2670} (\bibinfo {year} {2020})}\BibitemShut {NoStop}%
\bibitem [{\citenamefont {Ma}\ \emph {et~al.}(2015{\natexlab{a}})\citenamefont {Ma}, \citenamefont {Yang}, \citenamefont {Zhao},\ and\ \citenamefont {Wu}}]{Ma2015}%
  \BibitemOpen
  \bibfield  {author} {\bibinfo {author} {\bibfnamefont {F.}~\bibnamefont {Ma}}, \bibinfo {author} {\bibfnamefont {X.}~\bibnamefont {Yang}}, \bibinfo {author} {\bibfnamefont {H.}~\bibnamefont {Zhao}},\ and\ \bibinfo {author} {\bibfnamefont {N.}~\bibnamefont {Wu}},\ }\bibfield  {title} {\bibinfo {title} {Inducing propulsion of colloidal dimers by breaking the symmetry in electrohydrodynamic flow},\ }\href {https://doi.org/10.1103/PhysRevLett.115.208302} {\bibfield  {journal} {\bibinfo  {journal} {Phys. Rev. Lett.}\ }\textbf {\bibinfo {volume} {115}},\ \bibinfo {pages} {208302} (\bibinfo {year} {2015}{\natexlab{a}})}\BibitemShut {NoStop}%
\bibitem [{\citenamefont {Ma}\ \emph {et~al.}(2015{\natexlab{b}})\citenamefont {Ma}, \citenamefont {Wang}, \citenamefont {Wu},\ and\ \citenamefont {Wu}}]{Ma2015PNAS}%
  \BibitemOpen
  \bibfield  {author} {\bibinfo {author} {\bibfnamefont {F.}~\bibnamefont {Ma}}, \bibinfo {author} {\bibfnamefont {S.}~\bibnamefont {Wang}}, \bibinfo {author} {\bibfnamefont {D.~T.}\ \bibnamefont {Wu}},\ and\ \bibinfo {author} {\bibfnamefont {N.}~\bibnamefont {Wu}},\ }\bibfield  {title} {\bibinfo {title} {Electric-field–induced assembly and propulsion of chiral colloidal clusters},\ }\href {https://doi.org/10.1073/pnas.1502141112} {\bibfield  {journal} {\bibinfo  {journal} {PNAS}\ }\textbf {\bibinfo {volume} {112}},\ \bibinfo {pages} {6307} (\bibinfo {year} {2015}{\natexlab{b}})}\BibitemShut {NoStop}%
\bibitem [{\citenamefont {Yang}\ \emph {et~al.}(2019)\citenamefont {Yang}, \citenamefont {Johnson},\ and\ \citenamefont {Wu}}]{Yang2019}%
  \BibitemOpen
  \bibfield  {author} {\bibinfo {author} {\bibfnamefont {X.}~\bibnamefont {Yang}}, \bibinfo {author} {\bibfnamefont {S.}~\bibnamefont {Johnson}},\ and\ \bibinfo {author} {\bibfnamefont {N.}~\bibnamefont {Wu}},\ }\bibfield  {title} {\bibinfo {title} {The impact of stern-layer conductivity on the electrohydrodynamic flow around colloidal motors under an alternating current electric field},\ }\href {https://doi.org/10.1002/aisy.201900096} {\bibfield  {journal} {\bibinfo  {journal} {Adv. Intell. Syst.}\ }\textbf {\bibinfo {volume} {1}},\ \bibinfo {pages} {1900096} (\bibinfo {year} {2019})}\BibitemShut {NoStop}%
\bibitem [{\citenamefont {Allan}\ \emph {et~al.}(2018)\citenamefont {Allan}, \citenamefont {Caswell}, \citenamefont {Keim},\ and\ \citenamefont {van~der Wel}}]{Trackpy}%
  \BibitemOpen
  \bibfield  {author} {\bibinfo {author} {\bibfnamefont {D.~B.}\ \bibnamefont {Allan}}, \bibinfo {author} {\bibfnamefont {T.}~\bibnamefont {Caswell}}, \bibinfo {author} {\bibfnamefont {N.~C.}\ \bibnamefont {Keim}},\ and\ \bibinfo {author} {\bibfnamefont {C.~M.}\ \bibnamefont {van~der Wel}},\ }\href {https://doi.org/10.5281/zenodo.1226458} {\bibinfo {title} {trackpy: Trackpy v0.4.1}} (\bibinfo {year} {2018})\BibitemShut {NoStop}%
\bibitem [{\citenamefont {Yang}\ and\ \citenamefont {Wu}(2018)}]{Yang2017}%
  \BibitemOpen
  \bibfield  {author} {\bibinfo {author} {\bibfnamefont {X.}~\bibnamefont {Yang}}\ and\ \bibinfo {author} {\bibfnamefont {N.}~\bibnamefont {Wu}},\ }\bibfield  {title} {\bibinfo {title} {Change the collective behaviors of colloidal motors by tuning electrohydrodynamic flow at the subparticle level},\ }\href {https://doi.org/10.1021/acs.langmuir.7b02793} {\bibfield  {journal} {\bibinfo  {journal} {Langmuir}\ ,\ \bibinfo {pages} {952}} (\bibinfo {year} {2018})}\BibitemShut {NoStop}%
\bibitem [{\citenamefont {Janssen}(2019)}]{Janssen2019}%
  \BibitemOpen
  \bibfield  {author} {\bibinfo {author} {\bibfnamefont {L.~M.~C.}\ \bibnamefont {Janssen}},\ }\bibfield  {title} {\bibinfo {title} {Active glasses},\ }\href {https://doi.org/10.1088/1361-648X/ab3e90} {\bibfield  {journal} {\bibinfo  {journal} {J. Phys.: Condens. Matter}\ ,\ \bibinfo {pages} {503002}} (\bibinfo {year} {2019})}\BibitemShut {NoStop}%
\bibitem [{\citenamefont {Boltz}\ and\ \citenamefont {Ihle}(2024)}]{Boltz2024}%
  \BibitemOpen
  \bibfield  {author} {\bibinfo {author} {\bibfnamefont {H.-H.}\ \bibnamefont {Boltz}}\ and\ \bibinfo {author} {\bibfnamefont {T.}~\bibnamefont {Ihle}},\ }\bibfield  {title} {\bibinfo {title} {Hyperuniformity in deterministic anti-aligning active matter},\ }\bibfield  {journal} {\bibinfo  {journal} {arXiv:2402.19451}\ }\href {https://doi.org/https://doi.org/10.48550/arXiv.2402.19451} {https://doi.org/10.48550/arXiv.2402.19451} (\bibinfo {year} {2024})\BibitemShut {NoStop}%
\bibitem [{\citenamefont {Lei}\ and\ \citenamefont {Ni}(2019)}]{Lei2019}%
  \BibitemOpen
  \bibfield  {author} {\bibinfo {author} {\bibfnamefont {Q.-L.}\ \bibnamefont {Lei}}\ and\ \bibinfo {author} {\bibfnamefont {R.}~\bibnamefont {Ni}},\ }\bibfield  {title} {\bibinfo {title} {Hydrodynamics of random-organizing hyperuniform fluids},\ }\href {https://doi.org/10.1073/pnas.1911596116} {\bibfield  {journal} {\bibinfo  {journal} {PNAS}\ ,\ \bibinfo {pages} {22983}} (\bibinfo {year} {2019})}\BibitemShut {NoStop}%
\bibitem [{\citenamefont {Escaff}(2024{\natexlab{a}})}]{Escaff2024feb}%
  \BibitemOpen
  \bibfield  {author} {\bibinfo {author} {\bibfnamefont {D.}~\bibnamefont {Escaff}},\ }\bibfield  {title} {\bibinfo {title} {Anti-aligning interaction between active particles induces a finite wavelength instability: The dancing hexagons},\ }\href {https://doi.org/https://doi.org/10.1103/PhysRevE.109.024602} {\bibfield  {journal} {\bibinfo  {journal} {Phys. Rev. E}\ }\textbf {\bibinfo {volume} {109}},\ \bibinfo {pages} {024602} (\bibinfo {year} {2024}{\natexlab{a}})}\BibitemShut {NoStop}%
\bibitem [{\citenamefont {Escaff}(2024{\natexlab{b}})}]{Escaff2024aug}%
  \BibitemOpen
  \bibfield  {author} {\bibinfo {author} {\bibfnamefont {D.}~\bibnamefont {Escaff}},\ }\bibfield  {title} {\bibinfo {title} {Self-organization of anti-aligning active particles: Waving pattern formation and chaos},\ }\href {https://doi.org/https://doi.org/10.1103/PhysRevE.110.024603} {\bibfield  {journal} {\bibinfo  {journal} {Phys. Rev. E}\ }\textbf {\bibinfo {volume} {110}},\ \bibinfo {pages} {024603} (\bibinfo {year} {2024}{\natexlab{b}})}\BibitemShut {NoStop}%
\bibitem [{\citenamefont {Arold}\ and\ \citenamefont {Schmiedeberg}(2020)}]{Arold2020}%
  \BibitemOpen
  \bibfield  {author} {\bibinfo {author} {\bibfnamefont {D.}~\bibnamefont {Arold}}\ and\ \bibinfo {author} {\bibfnamefont {M.}~\bibnamefont {Schmiedeberg}},\ }\bibfield  {title} {\bibinfo {title} {Mean field approach of dynamical pattern formation in underdamped active matter with short-ranged alignment and distant anti-alignment interactions},\ }\href {https://doi.org/https://doi.org/10.1088/1361-648X/ab849b} {\bibfield  {journal} {\bibinfo  {journal} {J. Phys.: Condens. Matter}\ }\textbf {\bibinfo {volume} {32}},\ \bibinfo {pages} {315403} (\bibinfo {year} {2020})}\BibitemShut {NoStop}%
\bibitem [{\citenamefont {Zion}\ \emph {et~al.}(2023)\citenamefont {Zion}, \citenamefont {Fersula}, \citenamefont {Bredeche},\ and\ \citenamefont {Dauchot}}]{Zion2023}%
  \BibitemOpen
  \bibfield  {author} {\bibinfo {author} {\bibfnamefont {M.~Y.~B.}\ \bibnamefont {Zion}}, \bibinfo {author} {\bibfnamefont {J.}~\bibnamefont {Fersula}}, \bibinfo {author} {\bibfnamefont {N.}~\bibnamefont {Bredeche}},\ and\ \bibinfo {author} {\bibfnamefont {O.}~\bibnamefont {Dauchot}},\ }\bibfield  {title} {\bibinfo {title} {Morphological computation and decentralized learning in a swarm of sterically interacting robots},\ }\href {https://doi.org/https://doi.org/10.1126/scirobotics.abo6140} {\bibfield  {journal} {\bibinfo  {journal} {Sci. Robot.}\ }\textbf {\bibinfo {volume} {8}},\ \bibinfo {pages} {eabo6140} (\bibinfo {year} {2023})}\BibitemShut {NoStop}%
\bibitem [{\citenamefont {Mirhosseini}\ \emph {et~al.}(2022)\citenamefont {Mirhosseini}, \citenamefont {Zion}, \citenamefont {Dauchot},\ and\ \citenamefont {Bredeche}}]{Mirhosseini2022}%
  \BibitemOpen
  \bibfield  {author} {\bibinfo {author} {\bibfnamefont {Y.}~\bibnamefont {Mirhosseini}}, \bibinfo {author} {\bibfnamefont {M.~Y.~B.}\ \bibnamefont {Zion}}, \bibinfo {author} {\bibfnamefont {O.}~\bibnamefont {Dauchot}},\ and\ \bibinfo {author} {\bibfnamefont {N.}~\bibnamefont {Bredeche}},\ }\bibfield  {title} {\bibinfo {title} {Adaptive phototaxis of a swarm of mobile robots using positive and negative feedback self-alignment},\ }\href {https://doi.org/https://doi.org/10.1145/3512290.3528816} {\bibfield  {journal} {\bibinfo  {journal} {Proceedings of the Genetic and Evolutionary Computation Conference}\ }\textbf {\bibinfo {volume} {GECCO '22}},\ \bibinfo {pages} {104} (\bibinfo {year} {2022})}\BibitemShut {NoStop}%
\bibitem [{\citenamefont {Pethig}(2017)}]{Pethig2017}%
  \BibitemOpen
  \bibfield  {author} {\bibinfo {author} {\bibfnamefont {R.}~\bibnamefont {Pethig}},\ }\bibinfo {title} {The clausius--mossotti factor},\ in\ \href {https://doi.org/10.1002/9781118671443.ch6} {\emph {\bibinfo {booktitle} {Dielectrophoresis}}}\ (\bibinfo  {publisher} {John Wiley \& Sons, Hoboken, NJ},\ \bibinfo {year} {2017})\ Chap.~\bibinfo {chapter} {6}, pp.\ \bibinfo {pages} {119--144}\BibitemShut {NoStop}%
\bibitem [{\citenamefont {Shilov}\ \emph {et~al.}(2000)\citenamefont {Shilov}, \citenamefont {Delgado}, \citenamefont {Gonz{\'a}lez-Caballero}, \citenamefont {Horno}, \citenamefont {L{\'o}pez-Garc{\'\i}a},\ and\ \citenamefont {Grosse}}]{Shilov2000}%
  \BibitemOpen
  \bibfield  {author} {\bibinfo {author} {\bibfnamefont {V.}~\bibnamefont {Shilov}}, \bibinfo {author} {\bibfnamefont {A.}~\bibnamefont {Delgado}}, \bibinfo {author} {\bibfnamefont {F.}~\bibnamefont {Gonz{\'a}lez-Caballero}}, \bibinfo {author} {\bibfnamefont {J.}~\bibnamefont {Horno}}, \bibinfo {author} {\bibfnamefont {J.}~\bibnamefont {L{\'o}pez-Garc{\'\i}a}},\ and\ \bibinfo {author} {\bibfnamefont {C.}~\bibnamefont {Grosse}},\ }\bibfield  {title} {\bibinfo {title} {Polarization of the electrical double layer. time evolution after application of an electric field},\ }\href {https://doi.org/10.1006/jcis.2000.7152} {\bibfield  {journal} {\bibinfo  {journal} {J. Colloid Interf. Sci.}\ }\textbf {\bibinfo {volume} {232}},\ \bibinfo {pages} {141} (\bibinfo {year} {2000})}\BibitemShut {NoStop}%
\bibitem [{\citenamefont {Schmidt}\ \emph {et~al.}(2019)\citenamefont {Schmidt}, \citenamefont {Liebchen}, \citenamefont {Löwen},\ and\ \citenamefont {Volpe}}]{Schmidt2019}%
  \BibitemOpen
  \bibfield  {author} {\bibinfo {author} {\bibfnamefont {F.}~\bibnamefont {Schmidt}}, \bibinfo {author} {\bibfnamefont {B.}~\bibnamefont {Liebchen}}, \bibinfo {author} {\bibfnamefont {H.}~\bibnamefont {Löwen}},\ and\ \bibinfo {author} {\bibfnamefont {G.}~\bibnamefont {Volpe}},\ }\bibfield  {title} {\bibinfo {title} {Light-controlled assembly of active colloidal molecules},\ }\href {https://doi.org/10.1063/1.5079861} {\bibfield  {journal} {\bibinfo  {journal} {J. Chem. Phys.}\ }\textbf {\bibinfo {volume} {150}},\ \bibinfo {pages} {094905} (\bibinfo {year} {2019})}\BibitemShut {NoStop}%
\end{thebibliography}%

\clearpage

\subsection*{\large Appendix II: Supporting data}

\vspace{2cm}
\begin{figure*}[ht!]
    \centering
    \includegraphics[width=1\linewidth]{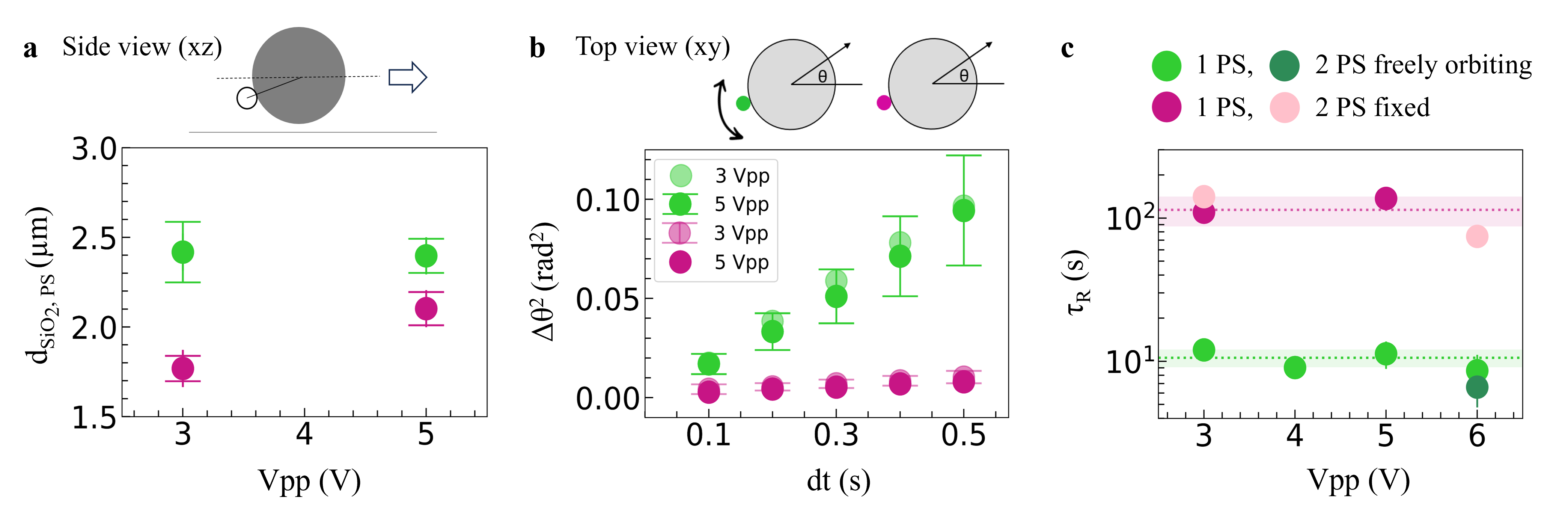}
    \caption{\textbf{a,} Measured center-to-center separation distance $d_{SiO_2, PS}$ between the SiO$_2$ and PS particle forming the self-reconfiguring (green) and pre-assembled (magenta) dimer indicating that molecules assume a different configuration with respect to the electrode as a function of field peak-to-peak amplitude $V_{pp}$. Self-reconfiguring molecules retain a constant configuration relative to the electrode. The schematic shows a side view of the molecule-electrode geometry. Errors denote standard deviations. \textbf{b,} Mean squared angular displacement MSAD curves of the core of self-reconfiguring (green) and pre-assembled (magenta) dimers. MSAD curves as a function of lag-time $dt$ are fitted to extract the rotational diffusion coefficients and corresponding timescales of reorientation as described in the manuscript. In the schematic, the angle $\theta$ represents the orientation of the velocity vector. Errors denote standard deviations. \textbf{c,} Timescale for reorientation $\tau_R$ as a function of $V_{pp}$ extracted from experiments on self-reconfiguring and pre-assembled dimers and trimers with one and two PS particles, respectively. Shaded areas denote the standard deviation. Standard errors are included but are not visible due to the logarithmic scale.}
    \label{fig:figS1_2}
\end{figure*} 

\clearpage

\vspace{4cm}
\begin{figure*}[ht!]
    \centering
    \includegraphics[width=0.8\linewidth]{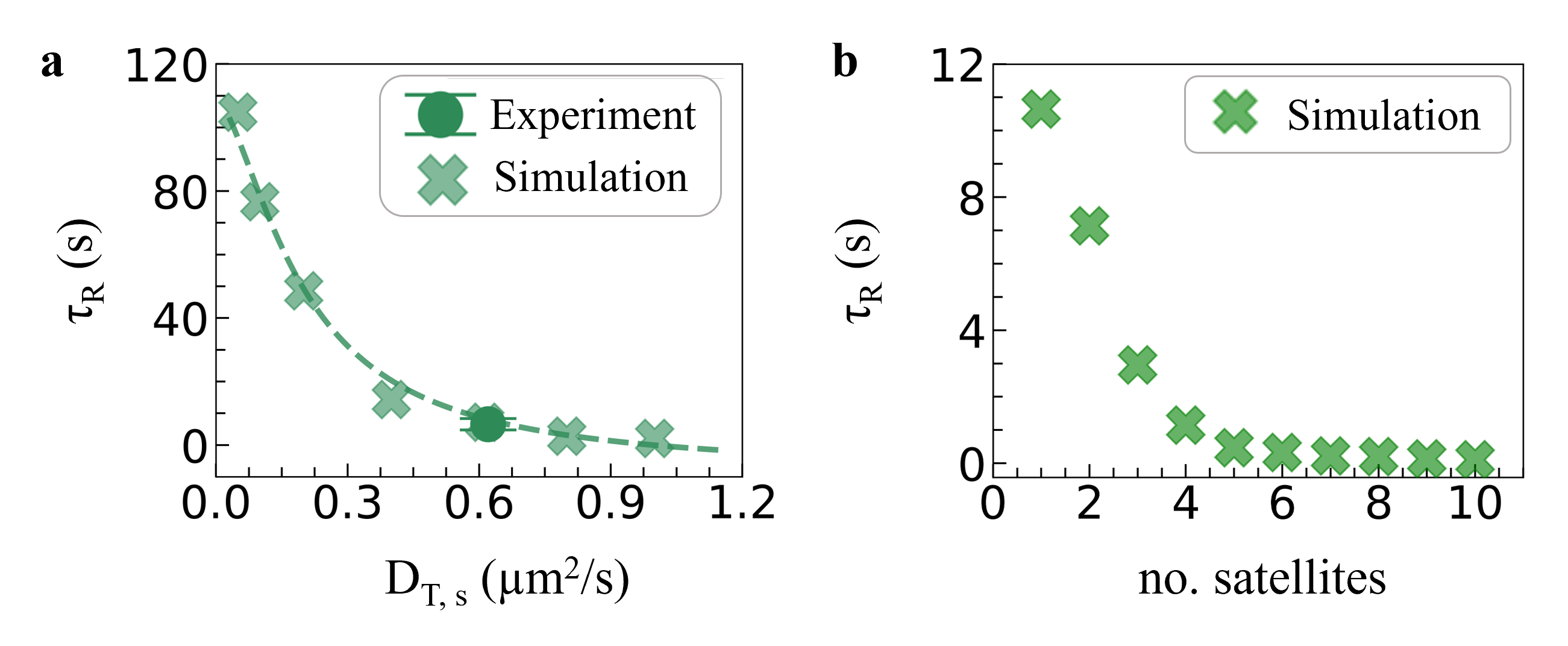}
    \caption{\textbf{a,} Timescale for reorientation $\tau_R$ as a function of satellite translational diffusivity $D_{T, s}$ for trimer molecules comprising a core and two satellite particles extracted from our simulations. Dashed line is a guide to the eye. \textbf{b,} Timescale for reorientation $\tau_R$ as a function of number of satellites extracted from simulations of self-reconfiguring molecules.}
    \label{fig:figS1_3}
\end{figure*} 

\begin{figure*}[h!]
    \centering
    \includegraphics[width=0.65\linewidth]{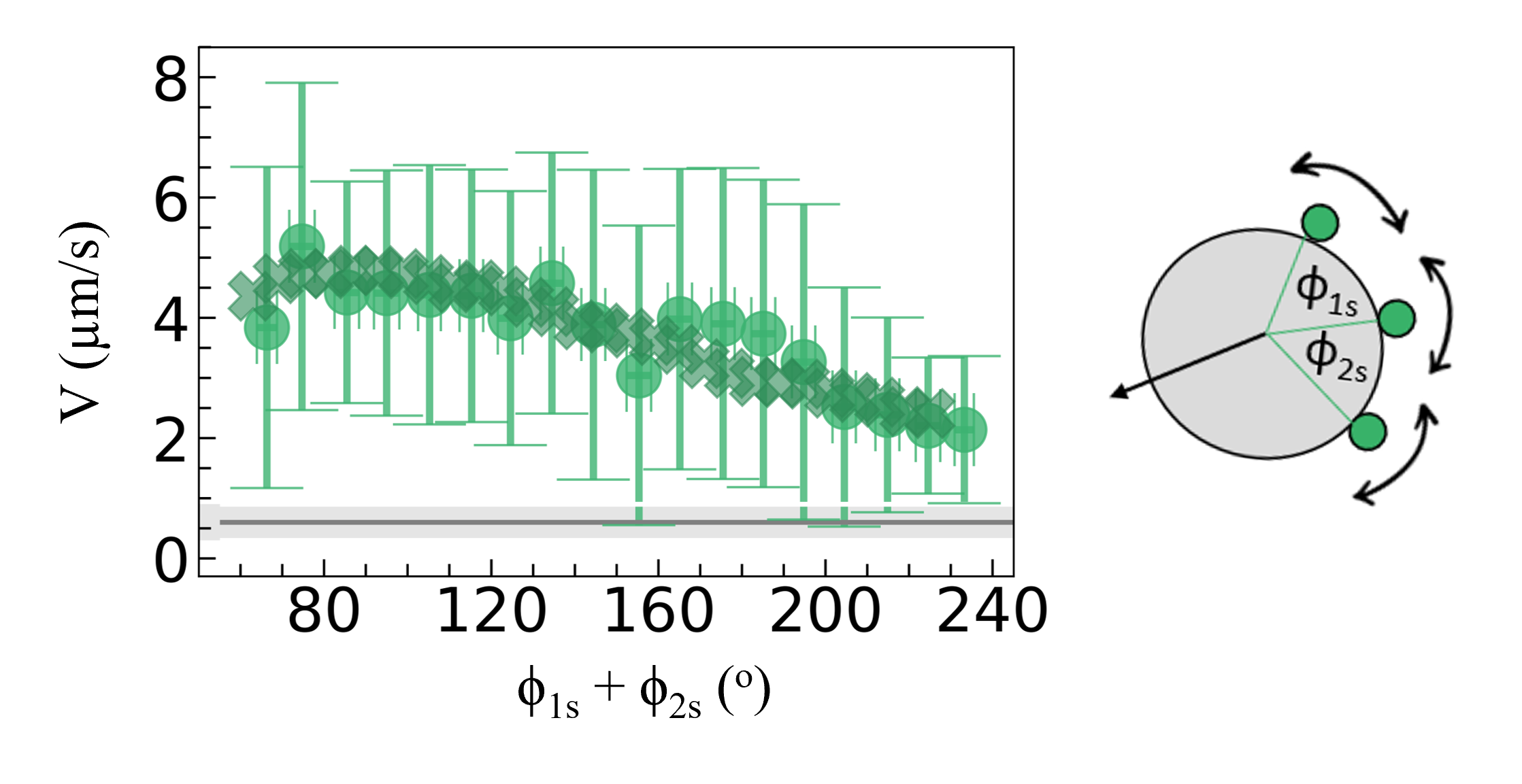}
    \caption{Instantaneous velocity $V$ of self-reconfiguring tetramers, consisting of a core and three satellite particles, decreases as a function of the instantaneous outer opening angle between satellites, given by the sum of the two smallest angles measured between the three satellites ($\phi_{1s}$ and $\phi_{2s}$, respectively), as illustrated in the schematic. Errors denote standard deviations. The gray line is the average speed and the shaded area the standard deviation of a single core particle obtained from its distribution of Brownian displacements at $dt$ = 0.1 s ($V_0$ = (0.6 $\pm$ 0.3) \textmu m/s). }
    \label{fig:figS1_3}
\end{figure*} 

\clearpage

\newpage

\vspace{30cm}
\begin{figure*}[ht!]
    \centering
    \includegraphics[width=1\linewidth]{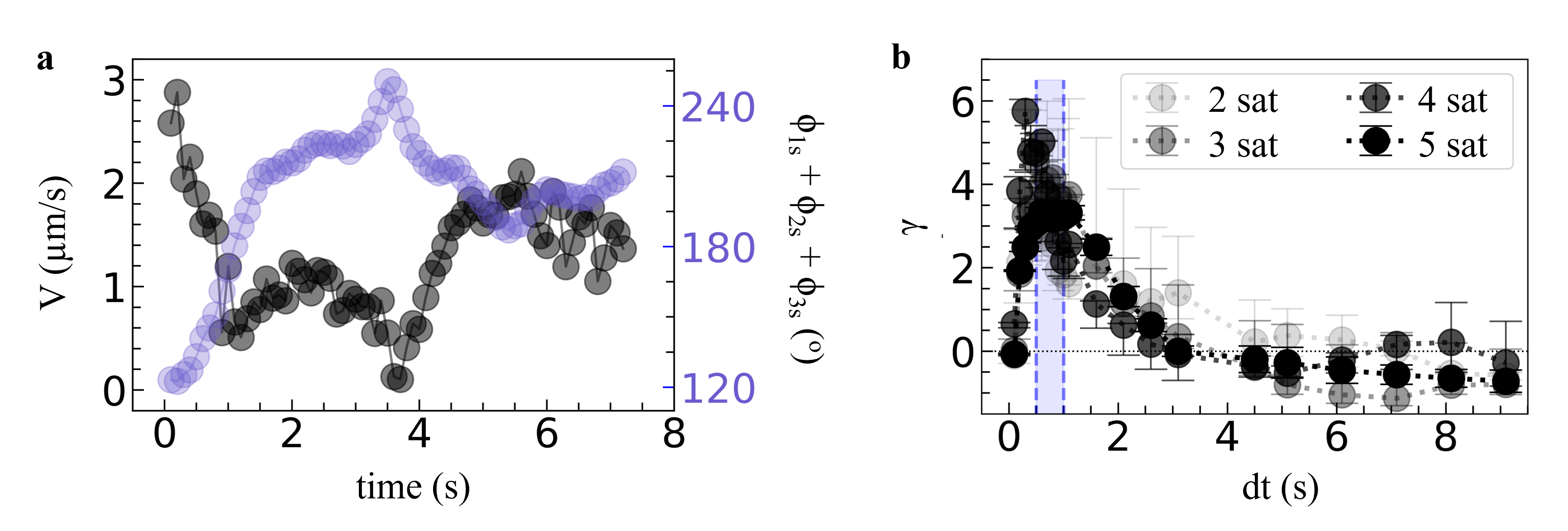}
    \caption{\textbf{a,} Speed $V$ of a self-reconfiguring pentamer with four satellites as a function of time as it approaches, ``collides'', and self-avoids its neighboring active molecule. Secondary $\psi$-axis shows the outer opening angle between satellites, i.e. the sum of the three smallest instantaneous angles between satellites ($\phi_{1s}$, $\phi_{2s}$, and $\phi_{3s}$, respectively) which follows the opposite trend, in agreement with the ``sum rule'' of the velocity. \textbf{b,} Kurtosis $\gamma$ of the angular displacement distributions of cores as a function of lag-time $dt$ for molecules with varying numbers of satellites that undergo consecutive ``collisions''. The height of the peak of the kurtosis increases with the number of collisions. The kurtosis peaks at $\approx$ 0.6 s commensurate with the duration of a collision (shaded region). Errors denote one standard deviation calculated from the 25 and 75\% quartiles. }
    \label{fig:figS1_3}
\end{figure*}

\end{document}